\documentclass[aps,prl,reprint,superscriptaddress]{revtex4-2}

\usepackage{amsmath} 
\usepackage{graphicx}
\usepackage{dcolumn}
\usepackage{bm}
\usepackage[utf8]{inputenc}
\usepackage[T1]{fontenc}
\usepackage{physics} 
\usepackage{color,soul}
\usepackage[colorlinks = true, linkcolor = blue, urlcolor  = blue, citecolor = blue, anchorcolor = blue]{hyperref}

\graphicspath{ {./images/} }

\newcommand{\Var}{\mathrm{Var}}
\newcommand{\cracklength}{l}
\newcommand{\crackspeed}{v}
\newcommand{\Gdyn}{G}
\newcommand{\Gstat}{G^{\mathrm{S}}}
\newcommand{\fracenergy}{\Gamma}
\newcommand{\homfracenergy}{\bar \Gamma}

\newcommand{\cs}{c_\mathrm{s}}
\newcommand{\cR}{c_\mathrm{R}}
\newcommand{\K}{K}

\newcommand{\vtr}{v_{\mathrm{c}}}
\newcommand{\lsys}{l_{\mathrm{sys}}}
\usepackage{moreverb} 

\begin{document}

\title{Effective toughness of heterogeneous materials with rate-dependent fracture energy}


\author{Gabriele Albertini}
\affiliation{Institute for Building Materials, ETH Zurich, Switzerland} 
\affiliation{School of Civil and Environmental Engineering, Cornell University, Ithaca NY, 14853, USA} 

\author{Mathias Lebihain}
\affiliation{Laboratoire Navier, ENPC/CNRS/IFSTTAR, France}
\affiliation{Institut Jean le Rond d’Alembert, Sorbonne Université/CNRS, France}

\author{Fran\c{c}ois Hild}
\affiliation{Laboratoire de Mécanique et Technologie (LMT), ENS Paris-Saclay/CNRS, France}

\author{Laurent Ponson}
\affiliation{Institut Jean le Rond d’Alembert, Sorbonne Université/CNRS, France}

\author{David S. Kammer}
\email{dkammer@ethz.ch}
\affiliation{Institute for Building Materials, ETH Zurich, Switzerland} 



\date{\today}

\begin{abstract}

We investigate dynamic fracture of heterogeneous materials experimentally by measuring displacement fields as a rupture propagates through a periodic array of obstacles of controlled fracture energy. Our measurements demonstrate the applicability of the classical equation of motion of cracks at a discontinuity of fracture energy: the crack speed jumps at the entrance and exit of an obstacle, as predicted by the crack-tip energy balance within the brittle fracture framework. The speed jump amplitude is governed by the fracture energy contrast and by the combination of rate-dependency of fracture energy and inertia of the medium, which allows the crack to cross a fracture energy discontinuity at constant energy release rate. This discontinuous dynamics and the rate-dependence cause higher effective toughness, which governs the coarse-grained behavior of these cracks.

  
\end{abstract}


\maketitle




Many biological materials, such as bone, nacre and tooth, have intricate microstructures which are responsible for remarkable macroscopic mechanical properties~\cite{
  ritchie_conflicts_2011,
  jackson_a._p._mechanical_1988
}.
Carefully designed microstructures combined with advances in micro-fabrication techniques allow for the development of new materials with unprecedented properties
\cite{florijn_programmable_2014, 
blees_graphene_2015,
bertoldi_negative_2010, 
silverberg_using_2014, 
siefert_bio-inspired_2019,
yin_impact-resistant_2019}. 
Understanding how to harness small-scale heterogeneities is, however, necessary to achieve the desired macroscopic properties.
For fracture properties, recent research focused either on disordered microstructures, where randomly located obstacles distort the crack front and cause toughening by collective pinning~\cite{gao_first-order_1989, roux_effective_2003, ponson_crack_2017,lebihain_effective_2020}, or on elastic heterogeneities, where compliant inclusions provide toughening by effectively reducing the energy flow into the crack tip~\cite{hossain_effective_2014,wang_cohesive_2017}.
However, a complete and fundamental theory for effective material resistance against fracture remains missing, and experimental observations, which are key for establishing such theoretical knowledge, are scarce. 

Theoretical fracture mechanics, based on the seminal work of Griffith~\cite{griffith_vi_1921,rice_thermodynamics_1978} states that a crack will propagate as soon as the released elastic energy per unit increment of crack length $\Gstat=-\partial_\cracklength \Omega$, where $\Omega$ is the elastic energy in the medium and $\cracklength$ the crack length, balances the local fracture energy $\fracenergy$ (\emph{i.e.}, the energy necessary for creating two unit surfaces). 
During dynamic crack propagation, the energy balance further includes inertia of the surrounding medium and possible rate-dependence of the fracture energy $\fracenergy(\crackspeed)$, where $\crackspeed=\dot l$ is the crack speed.
Using Linear Elastic Fracture Mechanics (LEFM) theory \cite{freund_dynamic_1990}, one can derive the equation of motion of a crack from this energy balance by assuming steady state crack propagation in an unbounded homogeneous domain. Under these circumstances the crack has no inertia (there is no term involving $\ddot l$ in the equation of motion) and its speed adapts abruptly to accommodate changes in fracture energy.
However, it remains unclear if these idealized conditions are valid at discontinuities within heterogeneous materials and how they affect the coarse-grained behavior of the crack during dynamic propagation.

In this Letter, we analyze these questions in depth through the experimental investigation of crack propagation in heterogeneous media with fracture energy discontinuities.  Usually, fracture mechanics experiments are based on global measurements, thus, only capture averaged quantities. 
In contrast, our experimental setup and simplified 2D geometry with periodic heterogeneities allows local measurements of the near-crack-tip fields, which support the uncovering of fundamental mechanisms.
While the elastic energy release rate is constant as the crack faces a fracture energy discontinuity, the speed at which the crack propagates is observed to vary discontinuously.
We study the amplitude of the speed jumps as the crack crosses the interface between regions of different fracture energy and show that it stems from the combination of rate-dependency of fracture energy and inertia of the medium. 
Rate-dependent effects result from the non-equilibrium nature of fracture problems and are prevailing in materials. 
Thus, rate-dependent fracture energy applies to a wide range of materials and has been observed, for instance, on rock~\cite{ponson_depinning_2009,atkinson_subcritical_1984}, glassy polymers~\cite{sharon_confirming_1999,livne_near-tip_2010, aagaard_near-source_2004,goldman_acquisition_2010,goldman_intrinsic_2012,
scheibert_brittle-quasibrittle_2010,vasudevan_adaptation_2021} and metals~\cite{rosakis_dynamic_1985}.
The discontinuous dynamics and the rate-dependent effects significantly affect the effective toughness of heterogeneous materials, as we will show with our experimental observations. 


\begin{figure} 
\includegraphics[width=3.375in]{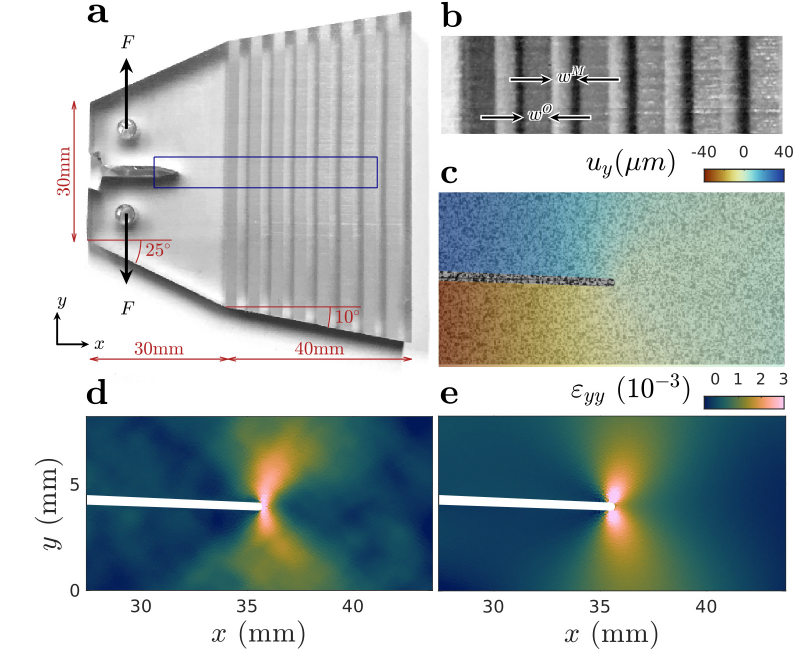}
\caption{\label{fig:method}
  (a)~Model heterogeneous material made of multi-material 3D-printed polymers in a tapered double cantilever beam geometry with applied forces $F$. The displacement field $\bm{u}=(u_x,u_y)$ is measured in the area within the blue box by digital image correlation.
  (b)~Closeup view shows two different materials in a periodic stripe geometry. The transparent material constitutes the matrix with width $w^M$ and the opaque (darker) areas are obstacles of higher fracture energy ${\fracenergy^O/\fracenergy^M\approx1.3}$ with width $w^O$.
  (c)~Closeup of crack tip at ${\cracklength\approx35\text{mm}}$ and ${\crackspeed\approx50\text{m/s}}$. The crack interface is slightly visible running from left to center. A random speckle pattern is applied onto the surface, which is compared to its reference pre-cracked configuration to find $\bm{u}$.
  (d)~Infinitesimal strain  ${\varepsilon_{yy}=\partial_y{u_y}}$ found by differentiating $\bm{u}$. 
  Approaching the crack tip, $\varepsilon_{yy}$ diverges.
  (e)~$\varepsilon_{yy}$ assuming the Williams eigenfunctions as basis for $\bm{u}$.}
\end{figure}

Our experimental setup (see FIG.~\ref{fig:method}a) consists of a tapered double cantilever beam, made of multi-material 3D-printed polymers (Stratasys Objet260 Connex3), a high-speed camera (Phantom v2511) and an  electromechanical testing machine (Shimadzu AG-X Plus). 
The matrix material is VeroClear with static fracture energy  ${\fracenergy^M_0\approx 80~\text{J/m}^2}$ and Young's modulus $E^M\approx 2.8\pm 0.2~\text{GPa}$.
The obstacle material is VeroWhite-DurusWhite ($\fracenergy_0^O\approx 106~\text{J/m}^2$, $E^O\approx 1.9\pm 0.2~\text{GPa}$), which is  tougher and more compliant.
We prescribe a constant crack mouth displacement rate ${\dot\delta\approx25\text{mm/s}}$. 
Hence, the elastic energy in the system is gradually increased, until a planar crack initiates from a pre-existing notch.
The elastic energy release rate at initiation is proportional to the bluntness of the notch, which we can tune to explore a range of initial crack speeds from moderate up to ${350~\text{m/s} \approx 0.4 \cR}$, where $\cR\approx800\text{m/s}$ is the Rayleigh wave speed. 
The crack propagates then dynamically through a series of periodic obstacles (see FIG.~\ref{fig:method}b).
During crack propagation no additional energy is added to the system ($\delta$ is constant) and the tapered geometry causes exponentially decaying released elastic energy  ${\Gstat\sim \delta^2 e^{-l/\lsys}}$, where $\lsys\approx 17.5~\text{mm}$ is a structural length scale directly related to the sample size~\cite{grabois_validation_2018}. 
Thus, the crack speed gradually decreases on average.
All properties are constant through the sample thickness and the overall behavior is {quasi-2D}.
We analyze the crack dynamics by measuring the near-tip displacement field $\bm{u}$ using Digital Image Correlation.
We apply a random speckle pattern (see FIG.~\ref{fig:method}c) onto the surface of the specimen using aerosol paint. The temporal evolution of the speckle is tracked using high speed photography at 250,000 fps.
The auto-correlation length of the pattern corresponds to 4-6 pixels, where the pixel size is $\approx45\mu\text{m}$.
$\bm{u}$ (see color in FIG.~\ref{fig:method}c) is found by minimizing the difference between the pattern at a given time $t$ mapped back to its pre-crack configuration~\cite{SM}.
The resulting infinitesimal strain field $\varepsilon_{yy}$ is depicted in FIG.~\ref{fig:method}d.
An alternative approach (see FIG.~\ref{fig:method}e) is the Integrated Digital Image Correlation (IDIC)~\cite{roux_stress_2006,grabois_validation_2018}, which assumes the analytical solution for a singular crack in an infinite elastic medium -- the Williams eigenfunctions expansion~\cite{williams_stress_1956} -- as basis for $\bm u$~\cite{SM}.
The first term of the series has singular strains at the crack tip ${\varepsilon_{ij} \sim 1/\sqrt{r}}$, where $r$ is the distance from the tip and its amplitude is related to the stress intensity factor $\K$.
Note that for both methods the amplitude of $\varepsilon$ is similar. IDIC has the advantages of precisely determining the crack tip position $l$ and directly computing $\K$, from which, one can find the dynamic energy release rate ${\Gdyn=\frac{\K^2}{E}A(v)}$ that provides a measure of the fracture energy $\fracenergy$ at the crack tip~\cite{freund_dynamic_1990,svetlizky_classical_2014,SM}.  The effects of elastic heterogeneity are minor in our setup, but give rise to an interaction between the size of the $K$-dominant region ($r \lesssim 5\text{mm}$) with the size of the heterogeneity and are discussed in~\cite{SM}.


\begin{figure} 
  \includegraphics[width=3.375in]{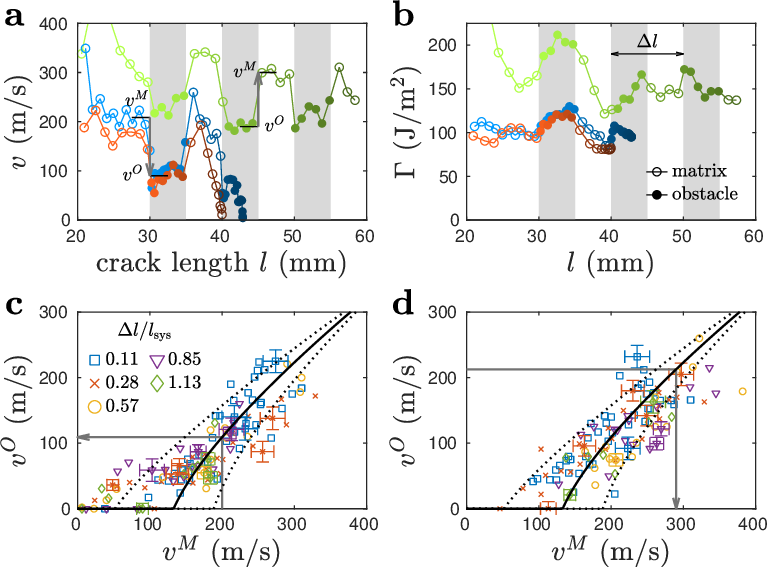}
  \caption{\label{fig:example_speed_jumps}
  (a,b)~Experimental results for three specimens with $\Delta l/\lsys=0.57$.
    (a)~$v$ undergoes abrupt deceleration ($\cracklength=\{30,40,50\}$mm) and acceleration ($\cracklength=\{35,45,55\}$mm) when the crack front is trapped and untrapped, respectively, at the interface.
    (b)~Discontinuities in $\fracenergy$ occur at trapping and untrapping with higher values within the obstacle.
    (c)~Trapping: speed prior to entering the obstacle $v^M$ is plotted vs. speed immediately after $v^O$. When the approaching velocity ${v^M<\vtr\approx 130~\text{m/s}}$ the front arrests.
    (d)~Untrapping: speed after exiting the obstacle $v^M$ is plotted vs. speed immediately before exiting $v^O$.
    (c,d)~Solid black line is the theoretical model (\ref{eq:eqmot_jump}) with $\pm10\%$ variation in $\fracenergy$ 
    (dotted lines).
  }
\end{figure}

Typical experiments are illustrated in FIG.~\ref{fig:example_speed_jumps}a\&b.
The crack first propagates through the matrix material with propagation speed $\crackspeed$ being maximum immediately after initiation, then $\crackspeed$ gradually decreases as crack length increases.
$\crackspeed$ undergoes abrupt deceleration (acceleration) as the front enters (leaves) an obstacle.
Simultaneously, $\fracenergy$ also abruptly increases (decreases). However, the relative jumps of the dissipation rate are significantly smaller than the ones observed on crack speed.
We calculate the speed in the obstacle $v^O$ and matrix $v^M$ by selecting the mean speed over $12\mu\text{s}$ before and after the obstacle boundaries.
All speed jumps at material discontinuities were studied for a collection of $30$ experiments with different period  $\Delta l=w^O+w^M$ and constant obstacle density ${\beta=\frac{w^O}{w^O+w^M}=1/2}$.
Jumps as the crack enters (trapping) and leaves (untrapping) an obstacle are shown in FIG.~\ref{fig:example_speed_jumps}c\&d, respectively.
Results show that the crack dynamics at the matrix/obstacle interface is independent of obstacle width and is symmetric with respect to the direction of propagation, \textit{i.e.}, the jumps are the same for trapping and untrapping. 
This implies that the crack dynamics only depends on local fracture properties.


In order to understand the jumps and their effect on effective material properties, we analyze the fracture propagation with a crack-tip energy balance.
In our experiments, failure mechanisms occur at time scales 4 orders of magnitude smaller than the viscous relaxation time typical of the polymers used in this study~\cite{SM} so that an elastic response of the sample can be safely assumed.
Moreover, the failure mechanisms are too fast for a craze to develop~\cite{ravi-chandar_mechanics_1988}, making the fracture process essentially brittle.
Thus, we develop a theoretical model based on LEFM to interpret the experimental observations. 

As the crack advances, elastic energy $\Gstat$ is released from the specimen  and is in part dissipated as fracture energy $\fracenergy$ to create new surfaces and in part radiated away as elastic waves.
Analyzing the near-tip fields of a steady-state dynamic crack,~\citet{freund_dynamic_1990} showed that the energy release rate of a dynamic crack $\Gdyn(l,v)$ is related to the energy release rate for a corresponding static crack $\Gstat(l)$ by $g(v)$, a universal function of $v$.
The crack-tip energy balance provides the equation of motion for a crack~\cite{SM}
\begin{equation}
  \label{eq:eqmot}
  \fracenergy(v) = \Gstat(l)g(v) \approx \Gstat(l)(1-v/\cR),
\end{equation}
which implies that within the framework of LEFM, a sub-Rayleigh crack in an infinite medium has no inertia and $v$ adjusts instantaneously to fluctuations in $\fracenergy$ or $\Gstat$~\cite{SM}.
Note that for rate-dependent materials, the fracture energy $\fracenergy(v)$ is not constant.

\begin{figure}
\includegraphics[width=3.375in]{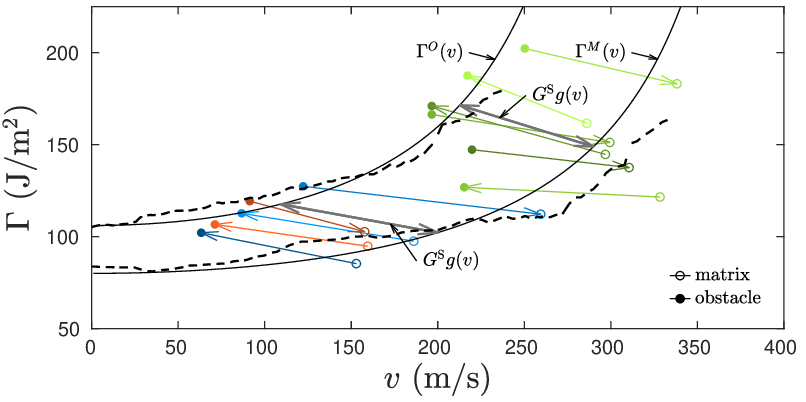}
\caption{\label{fig:speed_jumps_explained}
    Experimental results for the same specimens shown in FIG.~\ref{fig:example_speed_jumps}a\&b -- with same color-code. Data points represent crack speed and fracture energy at the moment of transition of material property.
    $\fracenergy(v)$ is separated in two distinct clusters corresponding to the matrix and obstacle material.
    Black dashed lines are the average fracture energy measurements based on 30 heterogeneous and 10 homogeneous samples~\cite{SM}.
    Solid black lines are the rate-dependent fracture energy law~\cite{scheibert_brittle-quasibrittle_2010} for the obstacle $\fracenergy^O(v)$ and matrix $\fracenergy^M(v)$ materials.
    The transition from one branch to the other is described by $\Gstat(l)g(v)$ -- the equation of the gray arrows  (\ref{eq:eqmot}).
  }
\end{figure}

We analyze the rate-dependence of the matrix and obstacle material by independently plotting $\fracenergy$ vs. $v$ (see averaged data as dashed line in FIG.~\ref{fig:speed_jumps_explained} or full data in FIG.~S3 of~\cite{SM}). We observe that our measurements are in good agreement with a model~\cite{scheibert_brittle-quasibrittle_2010} (solid line in FIG.~\ref{fig:speed_jumps_explained}) that considers the actual dissipative mechanism taking place within the process zone.
Within the matrix or obstacle material, the fracture energy follows this kinetic law. At the material boundaries, however, the rupture needs to jump from one kinetic law to the other.
The jump amplitude is governed by the equation of motion (\ref{eq:eqmot}). The jump trajectory in the $\fracenergy$-$\crackspeed$ space corresponds to the right-hand side of (\ref{eq:eqmot}), which, since $\Gstat(l)$ is constant across the boundary, corresponds to a diagonal line ${\Gstat g(\crackspeed)
}$ (arrows in FIG.~\ref{fig:speed_jumps_explained}).

Thus, at a discontinuity in material property the equation of motion of a crack becomes
\begin{equation}
  \label{eq:eqmot_jump}
  \Gstat=\fracenergy^M(v^M)/g(v^M)=\fracenergy^O(v^O)/g(v^O)~,
\end{equation}
which captures the experimentally observed velocity discontinuity at trapping and untrapping with no fitting parameter (see FIG.~\ref{fig:example_speed_jumps}c\&d).
Eq.~(\ref{eq:eqmot_jump}) cannot be solved explicitly. However, assuming a 
\emph{linear} rate-dependent fracture energy ${\fracenergy(v)\approx\fracenergy_0+\gamma v}$, for the purpose of discussion, the velocity jump becomes
\begin{equation}
    \label{eq:eqmot_jump_lin}
    v^M-v^O \approx \Delta \fracenergy_0 \frac{1-v^M/\cR}{\gamma +\fracenergy_0^M/\cR}~,
\end{equation}
where ${\Delta \fracenergy_0 = \fracenergy^O_0 - \fracenergy^M_0}$ is the jump in fracture energy.
This simple result highlights that
($i$)~the jump amplitude is the same for trapping 
and untrapping  (FIG.~\ref{fig:example_speed_jumps}c\&d) and
($ii$)~during trapping the velocity right after the interface is zero if $v^M$ is smaller than a critical incident velocity $\vtr$ below which the obstacle causes crack arrest 
\begin{equation}\label{eq:vthres}
\vtr \approx\Delta \fracenergy_0/\left(\gamma + \fracenergy_0^O/\cR\right)~.
\end{equation}
All these features are discernible from our experimental data and are captured  fairly well by the model.
Eq.~(\ref{eq:eqmot_jump_lin}) as well as a parameter study of (\ref{eq:eqmot_jump}) (see FIG.~S3 in~\cite{SM}) reveal that the speed jump and $\vtr$ are proportional to the toughness discontinuity ${\Delta \fracenergy_0}$. The latter is particularly noisy because of variations of fracture properties of both matrix and obstacle material, \emph{i.e.}, $\Var[\Delta \fracenergy_0]=\Var[\fracenergy_0^M]+\Var[\fracenergy_0^O]$, assuming $\fracenergy_0^O$ and $\fracenergy_0^M$ are uncorrelated.
In the limit of small rate-dependency ${\gamma \ll \fracenergy_0/\cR}$, inertia controls the speed jumps, that are then given by ${v^M-v^O \approx (\Delta \fracenergy_0 /\fracenergy_0^M)(\cR-v^M)}$ and the corresponding condition for crack arrest becomes ${v < \vtr \approx (\Delta \fracenergy_0/\fracenergy_0^O) \cR}$.
Conversely, in the limit of large rate-dependency $\gamma \gg \fracenergy_0/\cR$ and  quasi-static propagation $v\ll\cR$, inertia can be neglected and the speed jumps become constant ${v^M-v^O\approx \Delta \fracenergy_0/\gamma\equiv \vtr}$.


How does such a trapping/untrapping dynamics impact the effective fracture properties $\bar{\fracenergy}$ of heterogeneous materials?
We compute the homogenized fracture energy $\homfracenergy$  by integrating over an interval ${\Delta l}$ of uninterrupted crack propagation starting at $l_i$, the beginning of each matrix/obstacle period,
\begin{equation}
  \label{eq:hom_def}
  \homfracenergy(\bar v) =  \frac{1}{\Delta l} \int_{l_i}^{l_i+\Delta l} \fracenergy \left(v(\tilde l)\right) \dd \tilde l~.
\end{equation}
As  $\fracenergy$ in each phase depends on crack speed, $\homfracenergy$ depends on it too. Thus, we report $\bar{\fracenergy}$  as a function of the apparent crack velocity ${\bar v=\Delta l/\int_{l_i}^{l_i+\Delta l} v^{-1} \mathrm{d}l}$.

\begin{figure} 
\includegraphics[width=3.375in]{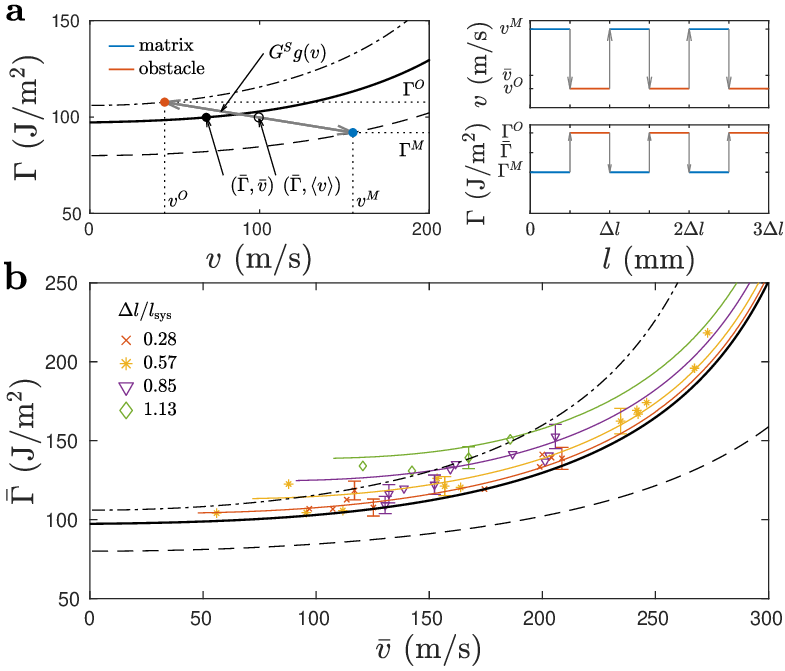}
\caption{\label{fig:homogenize}
  Homogenization of fracture energy ${\homfracenergy}$ vs.  average velocity, $\bar v$.  
  (a)~${\homfracenergy}$ assuming the scale separation condition  ${\Delta l \ll \lsys}$.
  Blue and red dots represent the state of the crack within the two materials, which are related by (\ref{eq:eqmot_jump}) depicted as a gray arrow.
  The black dot is the corresponding homogenized state ${(\bar \fracenergy, \bar v)}$ computed using (\ref{eq:hom_steady_f}) and (\ref{eq:hom_steady_v}).
  By varying $\Gstat$ one can derive the entire homogenized fracture energy law ${\bar \fracenergy (\bar v)}$ (black solid line in a\&b).
  (b)~${\homfracenergy( \bar v)}$, measured experimentally using (\ref{eq:hom_def}), is depicted as colored circles for a range of ${\Delta l \approx \lsys}$.
  Colored solid lines are the theoretical solution for ${\Delta l \approx \lsys}$ derived using (\ref{eq:hom_def}), (\ref{eq:eqmot})~\cite{SM}  (theory and experiment colors correspond).
  (a,b) Dash-dotted line and dashed line are $\fracenergy^O(v)$ and $\fracenergy^M(v)$ from FIG~\ref{fig:speed_jumps_explained}.
}
\end{figure}

First, we assume ${\Delta l \ll \lsys}$, \emph{i.e.}, a clear separation  between the micro-structural scale and the specimen scale. Hence, it is possible to define \emph{intrinsic} homogenized fracture properties, decoupled from the specimen size and the details of applied boundary conditions.
Under this assumption, $\Gstat$ remains constant during the entire crack propagation.
Thus,  $v$ and $\fracenergy$ are constant within each material phase (insets in FIG.~\ref{fig:homogenize}a), which allows us to calculate the dissipation rate from (\ref{eq:hom_def})
\begin{equation}
  \label{eq:hom_steady_f}
  \lim_{\Delta l/\lsys\rightarrow 0} \homfracenergy =\beta \fracenergy^O(v^O)+ (1-\beta) \fracenergy^M(v^M)
\end{equation}
and the apparent crack speed
\begin{equation}
  \label{eq:hom_steady_v}
\lim_{\Delta l/\lsys\rightarrow 0} \bar v = \left(\beta/v^O+(1-\beta)/v^M\right)^{-1},
\end{equation}
with ${\beta=1/2}$.
Note that (\ref{eq:hom_steady_v}) is a weighted harmonic mean, which is dominated by its lower argument, $v^O$, so $\bar{v}$ is effectively lower than the arithmetic mean ${(\langle v \rangle=\beta v^O + (1-\beta) v^M)}$.
As a result, the apparent kinetic law  $\bar \fracenergy(\bar v)$ is shifted ``horizontally'' towards lower speeds in comparison to ${\bar\fracenergy(\langle v \rangle)}$. This leads, in practice, to a resistance to failure $\bar{\fracenergy}$ \emph{larger} than the toughness spatial average ${\langle \fracenergy \rangle = \beta \fracenergy^{O}(\bar{v}) + (1-\beta) \fracenergy^M(\bar{v})}$, but lower than the obstacle toughness $\fracenergy^O$ predicted by rate-independent theory (see FIG.~\ref{fig:homogenize}).

However, when comparing the infinite system size prediction (\ref{eq:hom_steady_f}) and (\ref{eq:hom_steady_v}) to our
experimental measurements we observe higher effective toughness (see FIG.~\ref{fig:homogenize}b).
The interplay between the size of the heterogeneity $\Delta l$ and the structural length scale $\lsys$ makes homogenization of fracture properties particularly challenging.
The emerging effective toughness depends on the ratio ${\Delta l/\lsys}$, and (\ref{eq:hom_steady_f}) and (\ref{eq:hom_steady_v}) only represent a lower bound of $\homfracenergy(\bar v)$.
The larger ${\Delta l/\lsys}$, the higher ${\homfracenergy(\bar v)}$, which can even exceed $\fracenergy^O(\bar v)$ of the obstacle material.
This additional toughening, related to the structural problem with ${\Delta l\approx \lsys}$, is quantitatively captured by the theoretical solutions for ${\homfracenergy (\bar v)}$, which we derive from (\ref{eq:hom_def}) and (\ref{eq:eqmot}), assuming $\Gstat\sim e^{-l/\lsys}$.
Note that as we approach ${\Delta l \ll \lsys}$, the experimental toughness converges towards the theoretical one; and for ${\Delta l \gg \lsys}$ the rupture arrests before reaching $\Delta l$ required for homogenization of fracture properties. 

How do these observations translate to macroscopic measurements? While measurements from total elastic energy input (see FIG.~S5 in \cite{SM}) present increased toughness compared to the matrix material, they do not exceed the obstacle material. This is because the additional toughening observed at the small scale is a "horizontal shift" of the kinetic law. However, we observe that the macroscopic fracture energy is independent of $\Delta l$ and corresponds to the average of matrix and obstacle material, which validates (\ref{eq:hom_steady_f}). Furthermore, crack arrest, as described by (\ref{eq:vthres}), may play an important role in further increasing the macroscopic toughness. Even very thin obstacles may cause the crack to arrest, which raises interesting questions of practical importance for material design. How to design flaw insensitive materials, whose resistance to crack propagation -- or ability to prevent a crack to grow indefinitely -- is directly proportional to the obstacle toughness but independent of its size?
What are the strategies to translate this local toughening to the macro-scale and improve the mechanical integrity of structures through the use of damage-tolerant composites?


In summary, our study shows that the classical LEFM equation of motion of cracks quantitatively predicts crack dynamics at toughness discontinuities. The crack arrests if it is slower than a threshold speed that is primarily dependent on the toughness contrast and independent of the characteristic size of the microstructure (\emph{i.e.}, obstacle thickness), \textit{i.e.}, (\ref{eq:vthres}). When the crack penetrates the tougher/weaker obstacle, it reacts by instantaneously adapting its speed, which is mediated by the rate-dependent fracture energy combined with inertia, \textit{i.e.}, (\ref{eq:eqmot_jump_lin}). 
Finally, the heterogeneous material presents an increased effective (homogenized) toughness because of high fluctuations in crack speed between obstacles and matrix, and the rate-dependent nature of the fracture energy.
Direct experimental validation of (\ref{eq:eqmot_jump_lin}) and (\ref{eq:vthres}) is challenging due to limited temporal resolution and fluctuations in $\fracenergy$, but increased toughness contrast and focus on a single interface could provide a path to overcome these limitations.

\begin{acknowledgments}
       The authors thank Dr. Thiago Melo Grabois and Dr. Julien Scheibert for useful discussions.
\end{acknowledgments}


\begin{thebibliography}{29}%
\makeatletter
\providecommand \@ifxundefined [1]{%
 \@ifx{#1\undefined}
}%
\providecommand \@ifnum [1]{%
 \ifnum #1\expandafter \@firstoftwo
 \else \expandafter \@secondoftwo
 \fi
}%
\providecommand \@ifx [1]{%
 \ifx #1\expandafter \@firstoftwo
 \else \expandafter \@secondoftwo
 \fi
}%
\providecommand \natexlab [1]{#1}%
\providecommand \enquote  [1]{``#1''}%
\providecommand \bibnamefont  [1]{#1}%
\providecommand \bibfnamefont [1]{#1}%
\providecommand \citenamefont [1]{#1}%
\providecommand \href@noop [0]{\@secondoftwo}%
\providecommand \href [0]{\begingroup \@sanitize@url \@href}%
\providecommand \@href[1]{\@@startlink{#1}\@@href}%
\providecommand \@@href[1]{\endgroup#1\@@endlink}%
\providecommand \@sanitize@url [0]{\catcode `\\12\catcode `\$12\catcode
  `\&12\catcode `\#12\catcode `\^12\catcode `\_12\catcode `\%12\relax}%
\providecommand \@@startlink[1]{}%
\providecommand \@@endlink[0]{}%
\providecommand \url  [0]{\begingroup\@sanitize@url \@url }%
\providecommand \@url [1]{\endgroup\@href {#1}{\urlprefix }}%
\providecommand \urlprefix  [0]{URL }%
\providecommand \Eprint [0]{\href }%
\providecommand \doibase [0]{https://doi.org/}%
\providecommand \selectlanguage [0]{\@gobble}%
\providecommand \bibinfo  [0]{\@secondoftwo}%
\providecommand \bibfield  [0]{\@secondoftwo}%
\providecommand \translation [1]{[#1]}%
\providecommand \BibitemOpen [0]{}%
\providecommand \bibitemStop [0]{}%
\providecommand \bibitemNoStop [0]{.\EOS\space}%
\providecommand \EOS [0]{\spacefactor3000\relax}%
\providecommand \BibitemShut  [1]{\csname bibitem#1\endcsname}%
\let\auto@bib@innerbib\@empty
\bibitem [{\citenamefont {Ritchie}(2011)}]{ritchie_conflicts_2011}%
  \BibitemOpen
  \bibfield  {author} {\bibinfo {author} {\bibfnamefont {R.~O.}\ \bibnamefont
  {Ritchie}},\ }\href {https://doi.org/10.1038/nmat3115} {\bibfield  {journal}
  {\bibinfo  {journal} {Nature Materials}\ }\textbf {\bibinfo {volume} {10}},\
  \bibinfo {pages} {817} (\bibinfo {year} {2011})}\BibitemShut {NoStop}%
\bibitem [{\citenamefont {{Jackson A. P.}}\ \emph {et~al.}(1988)\citenamefont
  {{Jackson A. P.}}, \citenamefont {{Vincent Julian F. V.}}, \citenamefont
  {{Turner R. M.}},\ and\ \citenamefont {{Alexander Robert
  Mcneill}}}]{jackson_a._p._mechanical_1988}%
  \BibitemOpen
  \bibfield  {author} {\bibinfo {author} {\bibnamefont {{Jackson A. P.}}},
  \bibinfo {author} {\bibnamefont {{Vincent Julian F. V.}}}, \bibinfo {author}
  {\bibnamefont {{Turner R. M.}}},\ and\ \bibinfo {author} {\bibnamefont
  {{Alexander Robert Mcneill}}},\ }\href
  {https://doi.org/10.1098/rspb.1988.0056} {\bibfield  {journal} {\bibinfo
  {journal} {Proceedings of the Royal Society of London. Series B. Biological
  Sciences}\ }\textbf {\bibinfo {volume} {234}},\ \bibinfo {pages} {415}
  (\bibinfo {year} {1988})}\BibitemShut {NoStop}%
\bibitem [{\citenamefont {Florijn}\ \emph {et~al.}(2014)\citenamefont
  {Florijn}, \citenamefont {Coulais},\ and\ \citenamefont {van
  Hecke}}]{florijn_programmable_2014}%
  \BibitemOpen
  \bibfield  {author} {\bibinfo {author} {\bibfnamefont {B.}~\bibnamefont
  {Florijn}}, \bibinfo {author} {\bibfnamefont {C.}~\bibnamefont {Coulais}},\
  and\ \bibinfo {author} {\bibfnamefont {M.}~\bibnamefont {van Hecke}},\ }\href
  {https://doi.org/10.1103/PhysRevLett.113.175503} {\bibfield  {journal}
  {\bibinfo  {journal} {Physical Review Letters}\ }\textbf {\bibinfo {volume}
  {113}},\ \bibinfo {pages} {175503} (\bibinfo {year} {2014})}\BibitemShut
  {NoStop}%
\bibitem [{\citenamefont {Blees}\ \emph {et~al.}(2015)\citenamefont {Blees},
  \citenamefont {Barnard}, \citenamefont {Rose}, \citenamefont {Roberts},
  \citenamefont {McGill}, \citenamefont {Huang}, \citenamefont {Ruyack},
  \citenamefont {Kevek}, \citenamefont {Kobrin}, \citenamefont {Muller},\ and\
  \citenamefont {McEuen}}]{blees_graphene_2015}%
  \BibitemOpen
  \bibfield  {author} {\bibinfo {author} {\bibfnamefont {M.~K.}\ \bibnamefont
  {Blees}}, \bibinfo {author} {\bibfnamefont {A.~W.}\ \bibnamefont {Barnard}},
  \bibinfo {author} {\bibfnamefont {P.~A.}\ \bibnamefont {Rose}}, \bibinfo
  {author} {\bibfnamefont {S.~P.}\ \bibnamefont {Roberts}}, \bibinfo {author}
  {\bibfnamefont {K.~L.}\ \bibnamefont {McGill}}, \bibinfo {author}
  {\bibfnamefont {P.~Y.}\ \bibnamefont {Huang}}, \bibinfo {author}
  {\bibfnamefont {A.~R.}\ \bibnamefont {Ruyack}}, \bibinfo {author}
  {\bibfnamefont {J.~W.}\ \bibnamefont {Kevek}}, \bibinfo {author}
  {\bibfnamefont {B.}~\bibnamefont {Kobrin}}, \bibinfo {author} {\bibfnamefont
  {D.~A.}\ \bibnamefont {Muller}},\ and\ \bibinfo {author} {\bibfnamefont
  {P.~L.}\ \bibnamefont {McEuen}},\ }\href
  {https://doi.org/10.1038/nature14588} {\bibfield  {journal} {\bibinfo
  {journal} {Nature}\ }\textbf {\bibinfo {volume} {524}},\ \bibinfo {pages}
  {204} (\bibinfo {year} {2015})}\BibitemShut {NoStop}%
\bibitem [{\citenamefont {Bertoldi}\ \emph {et~al.}(2010)\citenamefont
  {Bertoldi}, \citenamefont {Reis}, \citenamefont {Willshaw},\ and\
  \citenamefont {Mullin}}]{bertoldi_negative_2010}%
  \BibitemOpen
  \bibfield  {author} {\bibinfo {author} {\bibfnamefont {K.}~\bibnamefont
  {Bertoldi}}, \bibinfo {author} {\bibfnamefont {P.~M.}\ \bibnamefont {Reis}},
  \bibinfo {author} {\bibfnamefont {S.}~\bibnamefont {Willshaw}},\ and\
  \bibinfo {author} {\bibfnamefont {T.}~\bibnamefont {Mullin}},\ }\href
  {https://doi.org/10.1002/adma.200901956} {\bibfield  {journal} {\bibinfo
  {journal} {Advanced Materials}\ }\textbf {\bibinfo {volume} {22}},\ \bibinfo
  {pages} {361} (\bibinfo {year} {2010})}\BibitemShut {NoStop}%
\bibitem [{\citenamefont {Silverberg}\ \emph {et~al.}(2014)\citenamefont
  {Silverberg}, \citenamefont {Evans}, \citenamefont {McLeod}, \citenamefont
  {Hayward}, \citenamefont {Hull}, \citenamefont {Santangelo},\ and\
  \citenamefont {Cohen}}]{silverberg_using_2014}%
  \BibitemOpen
  \bibfield  {author} {\bibinfo {author} {\bibfnamefont {J.~L.}\ \bibnamefont
  {Silverberg}}, \bibinfo {author} {\bibfnamefont {A.~A.}\ \bibnamefont
  {Evans}}, \bibinfo {author} {\bibfnamefont {L.}~\bibnamefont {McLeod}},
  \bibinfo {author} {\bibfnamefont {R.~C.}\ \bibnamefont {Hayward}}, \bibinfo
  {author} {\bibfnamefont {T.}~\bibnamefont {Hull}}, \bibinfo {author}
  {\bibfnamefont {C.~D.}\ \bibnamefont {Santangelo}},\ and\ \bibinfo {author}
  {\bibfnamefont {I.}~\bibnamefont {Cohen}},\ }\href
  {https://doi.org/10.1126/science.1252876} {\bibfield  {journal} {\bibinfo
  {journal} {Science}\ }\textbf {\bibinfo {volume} {345}},\ \bibinfo {pages}
  {647} (\bibinfo {year} {2014})}\BibitemShut {NoStop}%
\bibitem [{\citenamefont {Siéfert}\ \emph {et~al.}(2019)\citenamefont
  {Siéfert}, \citenamefont {Reyssat}, \citenamefont {Bico},\ and\
  \citenamefont {Roman}}]{siefert_bio-inspired_2019}%
  \BibitemOpen
  \bibfield  {author} {\bibinfo {author} {\bibfnamefont {E.}~\bibnamefont
  {Siéfert}}, \bibinfo {author} {\bibfnamefont {E.}~\bibnamefont {Reyssat}},
  \bibinfo {author} {\bibfnamefont {J.}~\bibnamefont {Bico}},\ and\ \bibinfo
  {author} {\bibfnamefont {B.}~\bibnamefont {Roman}},\ }\href
  {https://doi.org/10.1038/s41563-018-0219-x} {\bibfield  {journal} {\bibinfo
  {journal} {Nature Materials}\ }\textbf {\bibinfo {volume} {18}},\ \bibinfo
  {pages} {24} (\bibinfo {year} {2019})}\BibitemShut {NoStop}%
\bibitem [{\citenamefont {Yin}\ \emph {et~al.}(2019)\citenamefont {Yin},
  \citenamefont {Hannard},\ and\ \citenamefont
  {Barthelat}}]{yin_impact-resistant_2019}%
  \BibitemOpen
  \bibfield  {author} {\bibinfo {author} {\bibfnamefont {Z.}~\bibnamefont
  {Yin}}, \bibinfo {author} {\bibfnamefont {F.}~\bibnamefont {Hannard}},\ and\
  \bibinfo {author} {\bibfnamefont {F.}~\bibnamefont {Barthelat}},\ }\href
  {https://doi.org/10.1126/science.aaw8988} {\bibfield  {journal} {\bibinfo
  {journal} {Science}\ }\textbf {\bibinfo {volume} {364}},\ \bibinfo {pages}
  {1260} (\bibinfo {year} {2019})},\ \bibinfo {note} {publisher: American
  Association for the Advancement of Science Section: Report}\BibitemShut
  {NoStop}%
\bibitem [{\citenamefont {Gao}\ and\ \citenamefont
  {Rice}(1989)}]{gao_first-order_1989}%
  \BibitemOpen
  \bibfield  {author} {\bibinfo {author} {\bibfnamefont {H.}~\bibnamefont
  {Gao}}\ and\ \bibinfo {author} {\bibfnamefont {J.~R.}\ \bibnamefont {Rice}},\
  }\href {https://doi.org/10.1115/1.3176178} {\bibfield  {journal} {\bibinfo
  {journal} {Journal of Applied Mechanics}\ }\textbf {\bibinfo {volume} {56}},\
  \bibinfo {pages} {828} (\bibinfo {year} {1989})}\BibitemShut {NoStop}%
\bibitem [{\citenamefont {Roux}\ \emph {et~al.}(2003)\citenamefont {Roux},
  \citenamefont {Vandembroucq},\ and\ \citenamefont
  {Hild}}]{roux_effective_2003}%
  \BibitemOpen
  \bibfield  {author} {\bibinfo {author} {\bibfnamefont {S.}~\bibnamefont
  {Roux}}, \bibinfo {author} {\bibfnamefont {D.}~\bibnamefont {Vandembroucq}},\
  and\ \bibinfo {author} {\bibfnamefont {F.}~\bibnamefont {Hild}},\ }\href
  {https://doi.org/10.1016/S0997-7538(03)00078-0} {\bibfield  {journal}
  {\bibinfo  {journal} {European Journal of Mechanics - A/Solids}\ }\bibinfo
  {series} {General and plenary lectures from the 5th {EUROMECH} {Solid}
  {Mechanics} {Conference}},\ \textbf {\bibinfo {volume} {22}},\ \bibinfo
  {pages} {743} (\bibinfo {year} {2003})}\BibitemShut {NoStop}%
\bibitem [{\citenamefont {Ponson}\ and\ \citenamefont
  {Pindra}(2017)}]{ponson_crack_2017}%
  \BibitemOpen
  \bibfield  {author} {\bibinfo {author} {\bibfnamefont {L.}~\bibnamefont
  {Ponson}}\ and\ \bibinfo {author} {\bibfnamefont {N.}~\bibnamefont
  {Pindra}},\ }\href {https://doi.org/10.1103/PhysRevE.95.053004} {\bibfield
  {journal} {\bibinfo  {journal} {Physical Review E}\ }\textbf {\bibinfo
  {volume} {95}},\ \bibinfo {pages} {053004} (\bibinfo {year}
  {2017})}\BibitemShut {NoStop}%
\bibitem [{\citenamefont {Lebihain}\ \emph {et~al.}(2020)\citenamefont
  {Lebihain}, \citenamefont {Leblond},\ and\ \citenamefont
  {Ponson}}]{lebihain_effective_2020}%
  \BibitemOpen
  \bibfield  {author} {\bibinfo {author} {\bibfnamefont {M.}~\bibnamefont
  {Lebihain}}, \bibinfo {author} {\bibfnamefont {J.-B.}\ \bibnamefont
  {Leblond}},\ and\ \bibinfo {author} {\bibfnamefont {L.}~\bibnamefont
  {Ponson}},\ }\href {https://doi.org/10.1016/j.jmps.2020.103876} {\bibfield
  {journal} {\bibinfo  {journal} {Journal of the Mechanics and Physics of
  Solids}\ }\textbf {\bibinfo {volume} {137}},\ \bibinfo {pages} {103876}
  (\bibinfo {year} {2020})}\BibitemShut {NoStop}%
\bibitem [{\citenamefont {Hossain}\ \emph {et~al.}(2014)\citenamefont
  {Hossain}, \citenamefont {Hsueh}, \citenamefont {Bourdin},\ and\
  \citenamefont {Bhattacharya}}]{hossain_effective_2014}%
  \BibitemOpen
  \bibfield  {author} {\bibinfo {author} {\bibfnamefont {M.~Z.}\ \bibnamefont
  {Hossain}}, \bibinfo {author} {\bibfnamefont {C.~J.}\ \bibnamefont {Hsueh}},
  \bibinfo {author} {\bibfnamefont {B.}~\bibnamefont {Bourdin}},\ and\ \bibinfo
  {author} {\bibfnamefont {K.}~\bibnamefont {Bhattacharya}},\ }\href
  {https://doi.org/10.1016/j.jmps.2014.06.002} {\bibfield  {journal} {\bibinfo
  {journal} {Journal of the Mechanics and Physics of Solids}\ }\textbf
  {\bibinfo {volume} {71}},\ \bibinfo {pages} {15} (\bibinfo {year}
  {2014})}\BibitemShut {NoStop}%
\bibitem [{\citenamefont {Wang}\ and\ \citenamefont
  {Xia}(2017)}]{wang_cohesive_2017}%
  \BibitemOpen
  \bibfield  {author} {\bibinfo {author} {\bibfnamefont {N.}~\bibnamefont
  {Wang}}\ and\ \bibinfo {author} {\bibfnamefont {S.}~\bibnamefont {Xia}},\
  }\href {https://doi.org/10.1016/j.jmps.2016.09.004} {\bibfield  {journal}
  {\bibinfo  {journal} {Journal of the Mechanics and Physics of Solids}\
  }\textbf {\bibinfo {volume} {98}},\ \bibinfo {pages} {87} (\bibinfo {year}
  {2017})}\BibitemShut {NoStop}%
\bibitem [{\citenamefont {Griffith}\ and\ \citenamefont
  {Taylor}(1921)}]{griffith_vi_1921}%
  \BibitemOpen
  \bibfield  {author} {\bibinfo {author} {\bibfnamefont {A.~A.}\ \bibnamefont
  {Griffith}}\ and\ \bibinfo {author} {\bibfnamefont {G.~I.}\ \bibnamefont
  {Taylor}},\ }\href {https://doi.org/10.1098/rsta.1921.0006} {\bibfield
  {journal} {\bibinfo  {journal} {Philosophical Transactions of the Royal
  Society of London. Series A, Containing Papers of a Mathematical or Physical
  Character}\ }\textbf {\bibinfo {volume} {221}},\ \bibinfo {pages} {163}
  (\bibinfo {year} {1921})},\ \bibinfo {note} {publisher: Royal
  Society}\BibitemShut {NoStop}%
\bibitem [{\citenamefont {Rice}(1978)}]{rice_thermodynamics_1978}%
  \BibitemOpen
  \bibfield  {author} {\bibinfo {author} {\bibfnamefont {J.~R.}\ \bibnamefont
  {Rice}},\ }\href {https://doi.org/10.1016/0022-5096(78)90014-5} {\bibfield
  {journal} {\bibinfo  {journal} {Journal of the Mechanics and Physics of
  Solids}\ }\textbf {\bibinfo {volume} {26}},\ \bibinfo {pages} {61} (\bibinfo
  {year} {1978})}\BibitemShut {NoStop}%
\bibitem [{\citenamefont {Freund}(1990)}]{freund_dynamic_1990}%
  \BibitemOpen
  \bibfield  {author} {\bibinfo {author} {\bibfnamefont {L.~B.}\ \bibnamefont
  {Freund}},\ }\href {http://ebooks.cambridge.org/ref/id/CBO9780511546761}
  {\emph {\bibinfo {title} {Dynamic {Fracture} {Mechanics}}}}\ (\bibinfo
  {publisher} {Cambridge University Press},\ \bibinfo {address} {Cambridge},\
  \bibinfo {year} {1990})\BibitemShut {NoStop}%
\bibitem [{\citenamefont {Ponson}(2009)}]{ponson_depinning_2009}%
  \BibitemOpen
  \bibfield  {author} {\bibinfo {author} {\bibfnamefont {L.}~\bibnamefont
  {Ponson}},\ }\href {https://doi.org/10.1103/PhysRevLett.103.055501}
  {\bibfield  {journal} {\bibinfo  {journal} {Physical Review Letters}\
  }\textbf {\bibinfo {volume} {103}},\ \bibinfo {pages} {055501} (\bibinfo
  {year} {2009})}\BibitemShut {NoStop}%
\bibitem [{\citenamefont {Atkinson}(1984)}]{atkinson_subcritical_1984}%
  \BibitemOpen
  \bibfield  {author} {\bibinfo {author} {\bibfnamefont {B.~K.}\ \bibnamefont
  {Atkinson}},\ }\href {https://doi.org/10.1029/JB089iB06p04077} {\bibfield
  {journal} {\bibinfo  {journal} {Journal of Geophysical Research: Solid
  Earth}\ }\textbf {\bibinfo {volume} {89}},\ \bibinfo {pages} {4077} (\bibinfo
  {year} {1984})}\BibitemShut {NoStop}%
\bibitem [{\citenamefont {Sharon}\ and\ \citenamefont
  {Fineberg}(1999)}]{sharon_confirming_1999}%
  \BibitemOpen
  \bibfield  {author} {\bibinfo {author} {\bibfnamefont {E.}~\bibnamefont
  {Sharon}}\ and\ \bibinfo {author} {\bibfnamefont {J.}~\bibnamefont
  {Fineberg}},\ }\href {https://doi.org/10.1038/16891} {\bibfield  {journal}
  {\bibinfo  {journal} {Nature}\ }\textbf {\bibinfo {volume} {397}},\ \bibinfo
  {pages} {333} (\bibinfo {year} {1999})}\BibitemShut {NoStop}%
\bibitem [{\citenamefont {Livne}\ \emph {et~al.}(2010)\citenamefont {Livne},
  \citenamefont {Bouchbinder}, \citenamefont {Svetlizky},\ and\ \citenamefont
  {Fineberg}}]{livne_near-tip_2010}%
  \BibitemOpen
  \bibfield  {author} {\bibinfo {author} {\bibfnamefont {A.}~\bibnamefont
  {Livne}}, \bibinfo {author} {\bibfnamefont {E.}~\bibnamefont {Bouchbinder}},
  \bibinfo {author} {\bibfnamefont {I.}~\bibnamefont {Svetlizky}},\ and\
  \bibinfo {author} {\bibfnamefont {J.}~\bibnamefont {Fineberg}},\ }\href
  {https://doi.org/10.1126/science.1180476} {\bibfield  {journal} {\bibinfo
  {journal} {Science}\ }\textbf {\bibinfo {volume} {327}},\ \bibinfo {pages}
  {1359} (\bibinfo {year} {2010})}\BibitemShut {NoStop}%
\bibitem [{\citenamefont {Aagaard}\ and\ \citenamefont
  {Heaton}(2004)}]{aagaard_near-source_2004}%
  \BibitemOpen
  \bibfield  {author} {\bibinfo {author} {\bibfnamefont {B.~T.}\ \bibnamefont
  {Aagaard}}\ and\ \bibinfo {author} {\bibfnamefont {T.~H.}\ \bibnamefont
  {Heaton}},\ }\href {http://www.bssaonline.org/content/94/6/2064} {\bibfield
  {journal} {\bibinfo  {journal} {Bulletin of the Seismological Society of
  America}\ }\textbf {\bibinfo {volume} {94}},\ \bibinfo {pages} {2064}
  (\bibinfo {year} {2004})}\BibitemShut {NoStop}%
\bibitem [{\citenamefont {Goldman}\ \emph {et~al.}(2010)\citenamefont
  {Goldman}, \citenamefont {Livne},\ and\ \citenamefont
  {Fineberg}}]{goldman_acquisition_2010}%
  \BibitemOpen
  \bibfield  {author} {\bibinfo {author} {\bibfnamefont {T.}~\bibnamefont
  {Goldman}}, \bibinfo {author} {\bibfnamefont {A.}~\bibnamefont {Livne}},\
  and\ \bibinfo {author} {\bibfnamefont {J.}~\bibnamefont {Fineberg}},\ }\href
  {https://doi.org/10.1103/PhysRevLett.104.114301} {\bibfield  {journal}
  {\bibinfo  {journal} {Physical Review Letters}\ }\textbf {\bibinfo {volume}
  {104}},\ \bibinfo {pages} {114301} (\bibinfo {year} {2010})}\BibitemShut
  {NoStop}%
\bibitem [{\citenamefont {Goldman}\ \emph {et~al.}(2012)\citenamefont
  {Goldman}, \citenamefont {Harpaz}, \citenamefont {Bouchbinder},\ and\
  \citenamefont {Fineberg}}]{goldman_intrinsic_2012}%
  \BibitemOpen
  \bibfield  {author} {\bibinfo {author} {\bibfnamefont {T.}~\bibnamefont
  {Goldman}}, \bibinfo {author} {\bibfnamefont {R.}~\bibnamefont {Harpaz}},
  \bibinfo {author} {\bibfnamefont {E.}~\bibnamefont {Bouchbinder}},\ and\
  \bibinfo {author} {\bibfnamefont {J.}~\bibnamefont {Fineberg}},\ }\href
  {https://doi.org/10.1103/PhysRevLett.108.104303} {\bibfield  {journal}
  {\bibinfo  {journal} {Physical Review Letters}\ }\textbf {\bibinfo {volume}
  {108}},\ \bibinfo {pages} {104303} (\bibinfo {year} {2012})}\BibitemShut
  {NoStop}%
\bibitem [{\citenamefont {Scheibert}\ \emph {et~al.}(2010)\citenamefont
  {Scheibert}, \citenamefont {Guerra}, \citenamefont {Célarié}, \citenamefont
  {Dalmas},\ and\ \citenamefont
  {Bonamy}}]{scheibert_brittle-quasibrittle_2010}%
  \BibitemOpen
  \bibfield  {author} {\bibinfo {author} {\bibfnamefont {J.}~\bibnamefont
  {Scheibert}}, \bibinfo {author} {\bibfnamefont {C.}~\bibnamefont {Guerra}},
  \bibinfo {author} {\bibfnamefont {F.}~\bibnamefont {Célarié}}, \bibinfo
  {author} {\bibfnamefont {D.}~\bibnamefont {Dalmas}},\ and\ \bibinfo {author}
  {\bibfnamefont {D.}~\bibnamefont {Bonamy}},\ }\bibfield  {journal} {\bibinfo
  {journal} {Physical Review Letters}\ }\textbf {\bibinfo {volume} {104}},\
  \href {https://doi.org/10.1103/PhysRevLett.104.045501}
  {10.1103/PhysRevLett.104.045501} (\bibinfo {year} {2010})\BibitemShut
  {NoStop}%
\bibitem [{\citenamefont {Vasudevan}\ \emph {et~al.}(2021)\citenamefont
  {Vasudevan}, \citenamefont {Grabois}, \citenamefont {Cordeiro}, \citenamefont
  {Morel}, \citenamefont {Filho},\ and\ \citenamefont
  {Ponson}}]{vasudevan_adaptation_2021}%
  \BibitemOpen
  \bibfield  {author} {\bibinfo {author} {\bibfnamefont {A.}~\bibnamefont
  {Vasudevan}}, \bibinfo {author} {\bibfnamefont {T.~M.}\ \bibnamefont
  {Grabois}}, \bibinfo {author} {\bibfnamefont {G.~C.}\ \bibnamefont
  {Cordeiro}}, \bibinfo {author} {\bibfnamefont {S.}~\bibnamefont {Morel}},
  \bibinfo {author} {\bibfnamefont {R.~D.~T.}\ \bibnamefont {Filho}},\ and\
  \bibinfo {author} {\bibfnamefont {L.}~\bibnamefont {Ponson}},\ }\href
  {http://arxiv.org/abs/2101.04380} {\bibfield  {journal} {\bibinfo  {journal}
  {arXiv:2101.04380 [cond-mat]}\ } (\bibinfo {year} {2021})},\ \bibinfo {note}
  {arXiv: 2101.04380}\BibitemShut {NoStop}%
\bibitem [{\citenamefont {Rosakis}\ and\ \citenamefont
  {Zehnder}(1985)}]{rosakis_dynamic_1985}%
  \BibitemOpen
  \bibfield  {author} {\bibinfo {author} {\bibfnamefont {A.~J.}\ \bibnamefont
  {Rosakis}}\ and\ \bibinfo {author} {\bibfnamefont {A.~T.}\ \bibnamefont
  {Zehnder}},\ }\href {https://doi.org/10.1007/bf00017966} {\bibfield
  {journal} {\bibinfo  {journal} {International Journal of Fracture}\ }\textbf
  {\bibinfo {volume} {27}},\ \bibinfo {pages} {169} (\bibinfo {year} {1985})},\
  \bibinfo {note} {place: Dordrecht Publisher: Kluwer Academic
  Publishers}\BibitemShut {NoStop}%
\bibitem [{\citenamefont {Grabois}\ \emph {et~al.}(2018)\citenamefont
  {Grabois}, \citenamefont {Neggers}, \citenamefont {Ponson}, \citenamefont
  {Hild},\ and\ \citenamefont {Toledo~Filho}}]{grabois_validation_2018}%
  \BibitemOpen
  \bibfield  {author} {\bibinfo {author} {\bibfnamefont {T.~M.}\ \bibnamefont
  {Grabois}}, \bibinfo {author} {\bibfnamefont {J.}~\bibnamefont {Neggers}},
  \bibinfo {author} {\bibfnamefont {L.}~\bibnamefont {Ponson}}, \bibinfo
  {author} {\bibfnamefont {F.}~\bibnamefont {Hild}},\ and\ \bibinfo {author}
  {\bibfnamefont {R.~D.}\ \bibnamefont {Toledo~Filho}},\ }\href
  {https://doi.org/10.1016/j.engfracmech.2017.12.015} {\bibfield  {journal}
  {\bibinfo  {journal} {Engineering Fracture Mechanics}\ }\textbf {\bibinfo
  {volume} {191}},\ \bibinfo {pages} {311} (\bibinfo {year}
  {2018})}\BibitemShut {NoStop}%
\bibitem [{SM()}]{SM}%
  \BibitemOpen
  \href@noop {} {\bibinfo {title} {See supplemental material at [url will be
  inserted by publisher] for dynamic crack propagation theory, fracture energy
  measurements, equation of motion for a crack and integrated digital image
  correlation method based on the williams eigenfunctions. the equation of
  motion is applied to the trapping and untrapping and to the homogenization of
  fracture energy problems.}}\BibitemShut {Stop}%
\bibitem [{\citenamefont {Roux}\ and\ \citenamefont
  {Hild}(2006)}]{roux_stress_2006}%
  \BibitemOpen
  \bibfield  {author} {\bibinfo {author} {\bibfnamefont {S.}~\bibnamefont
  {Roux}}\ and\ \bibinfo {author} {\bibfnamefont {F.}~\bibnamefont {Hild}},\
  }\href {https://doi.org/10.1007/s10704-006-6631-2} {\bibfield  {journal}
  {\bibinfo  {journal} {International Journal of Fracture}\ }\textbf {\bibinfo
  {volume} {140}},\ \bibinfo {pages} {141} (\bibinfo {year}
  {2006})}\BibitemShut {NoStop}%
\bibitem [{\citenamefont {Williams}(1956)}]{williams_stress_1956}%
  \BibitemOpen
  \bibfield  {author} {\bibinfo {author} {\bibfnamefont {M.~L.}\ \bibnamefont
  {Williams}},\ }\href
  {http://resolver.caltech.edu/CaltechAUTHORS:20140729-122058948} {\bibfield
  {journal} {\bibinfo  {journal} {Journal of Applied Mechanics}\ }\textbf
  {\bibinfo {volume} {24}},\ \bibinfo {pages} {109} (\bibinfo {year}
  {1956})}\BibitemShut {NoStop}%
\bibitem [{\citenamefont {Svetlizky}\ and\ \citenamefont
  {Fineberg}(2014)}]{svetlizky_classical_2014}%
  \BibitemOpen
  \bibfield  {author} {\bibinfo {author} {\bibfnamefont {I.}~\bibnamefont
  {Svetlizky}}\ and\ \bibinfo {author} {\bibfnamefont {J.}~\bibnamefont
  {Fineberg}},\ }\href {https://doi.org/10.1038/nature13202} {\bibfield
  {journal} {\bibinfo  {journal} {Nature}\ }\textbf {\bibinfo {volume} {509}},\
  \bibinfo {pages} {205} (\bibinfo {year} {2014})}\BibitemShut {NoStop}%
\bibitem [{\citenamefont {Ravi-Chandar}\ and\ \citenamefont
  {Balzano}(1988)}]{ravi-chandar_mechanics_1988}%
  \BibitemOpen
  \bibfield  {author} {\bibinfo {author} {\bibfnamefont {K.}~\bibnamefont
  {Ravi-Chandar}}\ and\ \bibinfo {author} {\bibfnamefont {M.}~\bibnamefont
  {Balzano}},\ }\href {https://doi.org/10.1016/0013-7944(88)90161-0} {\bibfield
   {journal} {\bibinfo  {journal} {Engineering Fracture Mechanics}\ }\textbf
  {\bibinfo {volume} {30}},\ \bibinfo {pages} {713} (\bibinfo {year}
  {1988})}\BibitemShut {NoStop}%
\end{thebibliography}
\iftrue
\fi

\end{document}



\title{Supplemental Material \\
Effective toughness of heterogeneous materials with rate-dependent fracture energy }


\author{Gabriele Albertini}
\affiliation{Institute for Building Materials, ETH Zurich, Switzerland} 
\affiliation{School of Civil and Environmental Engineering, Cornell University, Ithaca NY, 14853, USA} 

\author{Mathias Lebihain}
\affiliation{Laboratoire Navier, ENPC/CNRS/IFSTTAR, France}
\affiliation{Institut Jean le Rond d’Alembert, Sorbonne Université/CNRS, France}

\author{Fran\c{c}ois Hild}
\affiliation{Laboratoire de Mécanique et Technologie (LMT), ENS Paris-Saclay/CNRS, France}

\author{Laurent Ponson}
\email{laurent.ponson@upmc.fr}
\affiliation{Institut Jean le Rond d’Alembert, Sorbonne Université/CNRS, France}

\author{David S. Kammer}
\email{dkammer@ethz.ch}
\affiliation{Institute for Building Materials, ETH Zurich, Switzerland} 



\date{\today}

\begin{abstract}
\end{abstract}


\maketitle


\appendix

\renewcommand{\thepage}{S\arabic{page}} 
\renewcommand{\thesection}{S\arabic{section}}  
\renewcommand{\thetable}{S\arabic{table}}  
\renewcommand{\thefigure}{S\arabic{figure}} 
\setcounter{figure}{0}
\setcounter{equation}{0}
\setcounter{page}{1}

\begin{table}[h]
\caption{\label{tab:variables}Description of variables used in the study}
\begin{ruledtabular}
\begin{tabular}{lll}
symbol & unit & description\\
\hline
$t$ & (s) & time\\
$u_i$ & (m) & displacement field, with $i=\{x,y\}$\\
$\varepsilon_{ij}$ & (-) & infinitesimal strain tensor\\
$\sigma_{ij}$ & (Pa) & stress tensor\\
$\sigma_Y$& (Pa) & yield stress \\
$E$ & (Pa)& Young's modulus\\
$\nu$ &(-)& Poisson's ratio\\
$\cl$ &(m/s) & dilatational wave speed\\
$\cs$ &(m/s) &  shear wave speed\\
$\cR$ &(m/s) &  Rayleigh wave speed\\
$\cracklength$ & (m) & crack length\\
$w^O$ & (m) & width of the obstacle stripe\\
$w^M$ & (m) & width of the matrix stripe\\
$\Delta l=w^O+w^M$ & (m) & period of the micro-structure\\
$\beta = w^O/\Delta l$ & (-) & obstacle density\\
$\crackspeed$ &(m/s)& crack speed\\
$\vtr$ &(m/s)& threshold speed causing crack arrest\\
$\lsys$ & (m)& structural length scale\\
$\K$ & (Pa $\sqrt{\text{m}}$) &stress intensity factor\\
$\K^\mathrm{S}$ & (Pa $\sqrt{\text{m}}$) & static stress intensity factor\\
$G$ & (J/m$^2$) & dynamic energy release rate\\
$\Gstat$ & (J/m$^2$)& static energy release rate\\
$\fracenergy$ & (J/m$^2$)& fracture energy\\
$\homfracenergy$ & (J/m$^2$)& effective fracture energy\\
$F$ & (N)& applied force at crack mouth\\
$\delta$ & (m) & crack mouth opening displacement\\
$\delta_m$ & (m) & loading apparatus displacement\\
$\lambda$ & (m/N) & compliance of the specimen\\
$\lambda_m$ & (m/N) & compliance of the loading apparatus\\
$b$ &(m)& specimen width\\
\end{tabular}
\end{ruledtabular}
\end{table}

\section{Experimental results}

\begin{figure}
    \centering
    \includegraphics[width=3.375in]{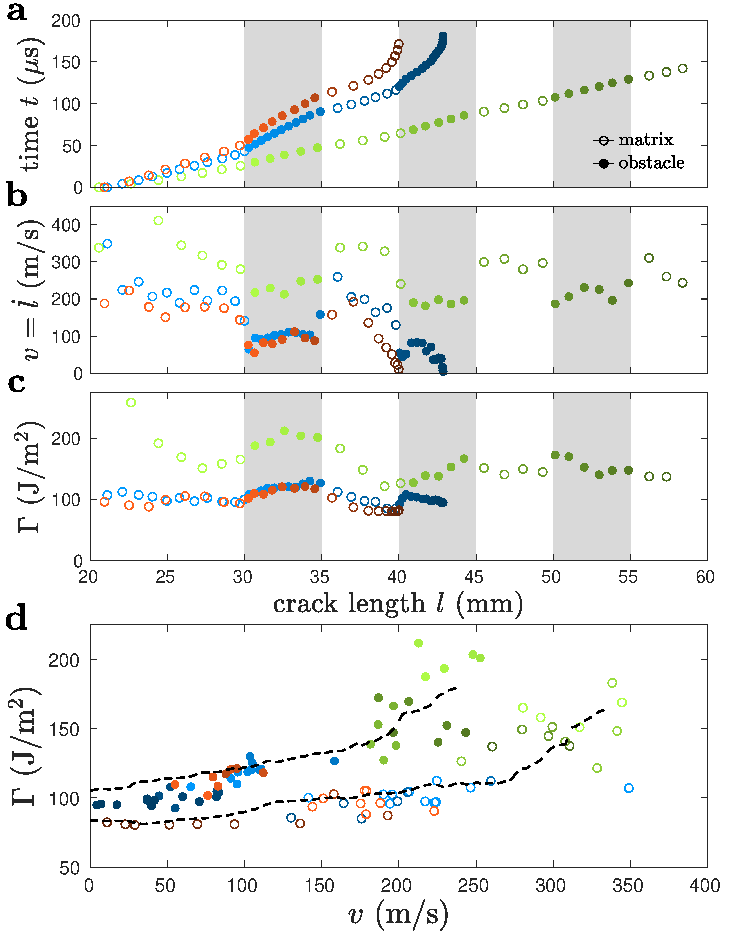}
    \caption{Supplemental experimental results to FIG.~2. Measurements of crack length vs. time (a) and corresponding crack speed (b) for the same samples as in FIG.~2 and 3.  
    (c)~Fracture energy $\fracenergy$ measured by Integrated Digital Image Correlation  vs. crack length. (d)~$\fracenergy(v)$ with average measurements  based on 30 heterogeneous and 10 homogeneous samples as black dashed lines.
    (a,b,c)~Gray areas represent obstacles.}
    \label{fig:crack_speed}
\end{figure}

Given the crack tip displacement versus time measurements $l(t)$, we compute the speed $v(t)$ by central differences
\begin{equation}
    v(l(t)+l(t+\Delta t)/2)=\frac{l(t)+l(t+\Delta t)}{\Delta t},
\end{equation}
where $\Delta t\approx 4\mu$s is the time interval between measurements.
FIG.~\ref{fig:crack_speed} shows the measured crack length, speed and fracture energy for selected experiments.

\section{Dynamic crack propagation}

\begin{figure}[h]
\includegraphics[width=3.375in]{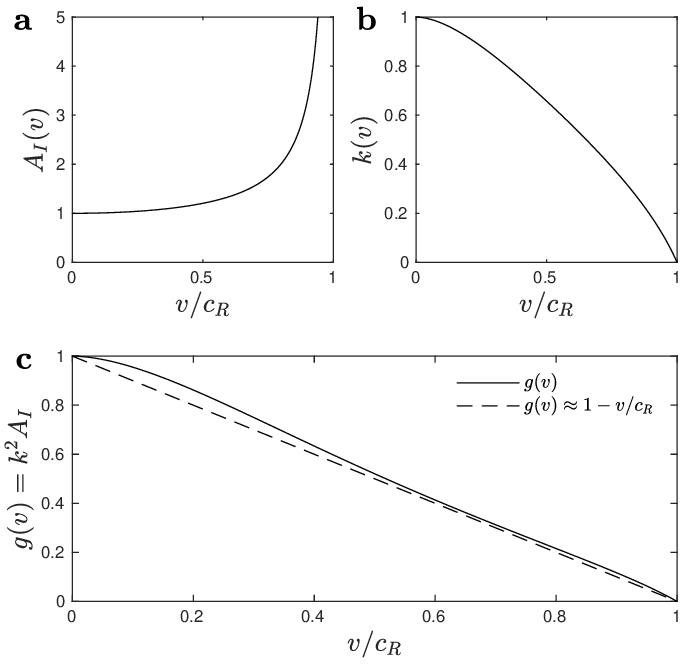}%
\caption{\label{fig:smlefm} Universal functions of dynamic fracture mechanics. (a)~Experiments are performed at velocities $\crackspeed<0.4\cR$ where the contribution of $A_I$ is small. (b, c)~Contribution of $k(v)$ is non-negligible for the considered $v$.}
\end{figure}

In our experiments, failure occurs during $\tau_{failure}=\lsys/v\sim=0.1~\text{ms}$ while the viscous relaxation time typical of the polymers used in this study is $\sim 1\text{s}$ so that an elastic response of the sample can be safely assumed.
The relaxation time is measured by applying a small crack mouth opening displacement such that the crack does not propagate and measuring the time over which the load decays.
In the present case, the boundaries are far enough from the crack tip so that reflected waves do not affect the dynamics and the infinite medium assumption holds.
A theoretical model based on Linear Elastic Fracture
Mechanics (LEFM) is thus developed to interpret the experimental
observations.
The dynamic energy release rate for plane stress configuration is then given by
\begin{equation}
  \label{eq:dynamicG}
G(\cracklength,\crackspeed,\loading) = \frac{1}{E}\K^2(\cracklength,\crackspeed,\loading)A_I(v) 
\end{equation}
where the function $A_I(v)$ is a universal function, in the sense that it is independent of applied loading $\loading$ or geometry, and $\K$ is the dynamic stress intensity factor \cite{freund_dynamic_1990}.
Assuming a semi-infinite crack in an unbounded linear elastic medium subjected to time independent loading, the dynamic stress intensity factor becomes
\begin{equation}
  \K(\cracklength,\crackspeed,\loading) = k(\crackspeed)\K^S(\cracklength,\loading) 
\end{equation}
where $k(\crackspeed)\approx (1- v/\cR)/\sqrt{1-v/\cl}$ is another universal function, $\cl$ is the dilatational wave speed and $\K^S(\cracklength,\loading)$ is the stress intensity factor for the equivalent static crack, which depends on geometry and applied loading \cite{freund_dynamic_1990}.
Thus, the dynamic energy release rate $G(\cracklength,\crackspeed,\loading)$ is related to the static energy release rate $\Gstat(\cracklength,\loading)$ by the universal function $g(v)$:
\begin{equation}\label{eq:dynamicG-Gs}
G(\cracklength,\crackspeed,\loading)=\Gstat(\cracklength,\loading) g(v),
\end{equation}
which can be approximated by $g(v)\approx 1-\crackspeed/\cR$ as illustrated in FIG.~\ref{fig:smlefm}.

\section{Fracture energy based on dynamic stress intensity factor measurements}

\begin{figure}
\includegraphics[width=3.375in]{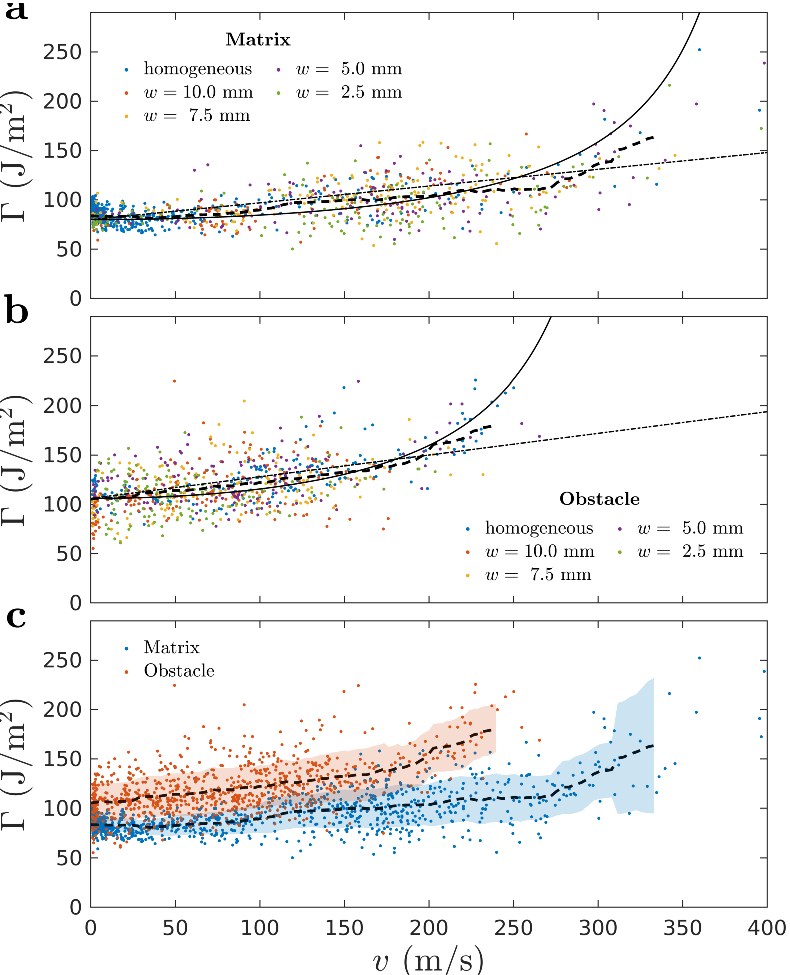}%
\caption{\label{fig:smcrc} Measured $\fracenergy(v)$ for the matrix (a) and obstacle (b) materials. Measurements on homogeneous samples are reported as well as for stripes geometry with different stripe widths. The dashed line is a moving average over an interval $\Delta v=25~\text{m/s}$. The solid line is the fracture energy model (\ref{eq:scheibert}). The dash-dotted line is the simplified fracture energy model (\ref{eq:simpfrac}).
(c)~Comparison of $\fracenergy(v)$ of both materials with error-band corresponding to standard deviation.}
\end{figure}

The stress intensity factor is measured from the displacement field, which means that the dynamic stress intensity factor $\K(\cracklength,\crackspeed,\loading)$ can be estimated. 
The dissipated fracture energy $\fracenergy(v)=\frac{\K^2}{E}A_I(v)$ is determined from a crack tip energy balance, where $A_I(v)$ accounts for the dynamic contribution of the energy release rate.
Rate-dependent processes occurring within the fracture process zone are accounted for by considering a speed dependent fracture energy model for speeds below the micro-branching instability \cite{scheibert_brittle-quasibrittle_2010}.
The fracture energy is related to the surface energy $\gamma^{\mathrm{s}}$ and the process zone size $l_{pz}(v)= \K(l,v)^2/a\sigma_Y^2=\K^S(l)^2k(v)^2/a\sigma_Y^2$
\begin{alignat*}{1}
  \fracenergy(v) &= \gamma^{\mathrm{s}} + \epsilon l_{pz}(v)\\
  &= \gamma^{\mathrm{s}} + \epsilon \frac{\K^S(l)^2k(v)^2}{a\sigma_Y^2}\\
  &= \gamma^{\mathrm{s}} + \epsilon \frac{\fracenergy(v)/A_I(v)E}{a\sigma_Y^2}
\end{alignat*}
which yields
\begin{alignat*}{1}
  \fracenergy(v) &= \frac{\gamma^{\mathrm{s}}}{1-\alpha/A_I(v)},
\end{alignat*}
where $\alpha = \epsilon E/(a \sigma_Y^2)$. Setting $\fracenergy_0=\gamma^{\mathrm{s}}/(1-\alpha)$ one finds
\begin{alignat}{1}
  \label{eq:scheibert}
  \fracenergy(v) = \fracenergy_0\frac{1-\alpha}{1-\alpha/A_I(v)},
\end{alignat}
where $\fracenergy_0$ and $\alpha$ are material dependent fitting parameters.
Postmortem fracture surfaces are smooth and microcrack branching from the main crack is absent.
All experiments have crack velocities lower than the critical velocity for the micro-branching instability to occur.
FIG.~\ref{fig:smcrc} shows the rate-dependent fracture energy for a collection of 30 experiments on homogeneous samples and stripe geometry.
From this study, it is found $\fracenergy^M_0=80~\text{J/m}^2$ for the matrix material and $\fracenergy^O_0=106~\text{J/m}^2$ for the obstacle, $\alpha =1.17$ as in Ref.~\cite{scheibert_brittle-quasibrittle_2010}. $A_I$ depends on the elastic properties of the material, which are $E^M\approx 2.8\pm 0.2~\text{GPa}$ for the matrix and $E^O\approx 1.9\pm 0.2~\text{GPa}$ for the obstacle,  $\nu\approx0.35$ and $\rho\approx1100~\text{kg/m}^3$ are identical for both materials. Assuming plane-stress condition one finds $\cR^M\approx900~\text{m/s}$ and $\cR^O\approx730~\text{m/s}$.

A simplified fracture energy model with linear dependency on $v$ is also introduced
\begin{equation}
  \label{eq:simpfrac}
  \Gamma(v) = \Gamma_0+\gamma v
\end{equation}
where $\gamma^O\approx0.22$ and $\gamma^M\approx0.17$.  

\section{Equation of motion of a crack}

The equation of motion of a crack is described by energy balance
\begin{equation}
  \label{eq:smeqmot}
  \fracenergy(v) = G(\cracklength,\crackspeed,\loading)
\end{equation}
where $\fracenergy(v)$ is the energy dissipated by a unit increment in crack length and $G$ is the dynamic energy release rate, given in equation~(\ref{eq:dynamicG-Gs}).
Note that $\crackspeed$ is first order in time meaning that the crack has no inertia.
The crack speed adjusts instantaneously to the speed dictated by $\fracenergy$ and $G$.

For a linear elastic 2D medium the strain energy is $\Omega= \frac{F^2}{2b}\lambda$, where $F$ is the applied force, $\lambda$ the compliance and $b=8\text{mm}$ is the specimen width. The static energy release rate becomes
\begin{equation}
  \label{eq:Gscmp}
  \Gstat=-\pdv{\Omega}{\cracklength}=\frac{F^2}{2b}\pdv{\lambda}{\cracklength}.
\end{equation}
For a displacement controlled system and accounting for the  compliance of the loading apparatus $\lambda_m$, the applied force becomes $F=\delta/(\lambda + \lambda_m)$, where $\delta$ is the prescribed displacement. The static energy release rate (\ref{eq:Gscmp}) remains unchanged but its rate of change is affected 
\begin{equation}
  \label{eq:dGsdl}
  \pdv{\Gstat}{\cracklength} = -\frac{F^2}{b} \frac{(\pdv*{\lambda}{\cracklength})^2}{\lambda+\lambda_m} + \frac{F^2}{2b}\pdv[2]{\lambda}{\cracklength}.
\end{equation}
Because the first term in equation~(\ref{eq:dGsdl}) is always negative, a nonzero $\lambda_m$ causes a larger $\pdv*{\Gstat}{\cracklength}$ compared to the case of an infinitely stiff loading stage $\lambda_m=0$. Thus, $\lambda_m$ is destabilizing the crack growth. However, $\pdv*{\Gstat}{\cracklength}<0$ keeps the crack velocities within the limits prescribed by the image acquisition setup.

The experiments were performed at a constant prescribed displacement rate $\dv{\delta}{t}=25~\text{mm/s}$. At typical propagation speed of $ \sim 100~\text{m/s}$, the crack breaks the specimen in $\sim 10^{-4}~\text{s}$. The typical loading time to reach the critical energy release rate for the crack to start propagating is $\sim 10^{-1}~\text{s}$, hence the change in $\delta$ during propagation is negligible $\sim 0.1 \%$.
An estimate $\lambda_m\approx 0.75~\mu\text{m/N}$ is based on geometry and elastic properties of the pins linking the specimen to the grips.
For comparison purposes, the specimen compliance is $\lambda \approx 0.7~\mu\text{m/N}$ at initiation ($l=20~\text{mm}$) and increases exponentially with crack length, reaching $\lambda \approx 25~\mu\text{m/N}$ at $l=60~\text{mm}$.

For the tapered double cantilever beam geometry, $\lambda(\cracklength)$ is computed by means of finite element simulations, solving the 2D elasticity equations. The geometry is chosen such that the compliance can be approximated by an exponential function
\begin{equation}
  \label{eq:compliance}
  \lambda=\frac{\lambda_0}{Eb}e^{\cracklength/\lsys},
\end{equation}
where $\lambda_0=3.45$ and $\lsys=17.5~\text{mm}$ depend on specimen geometry and $b=8~\text{mm}$ is the out-of-plane dimension, which are all kept constant during this study and are found by fitting equation~(\ref{eq:compliance}) to the finite element results.

We use Freund’s equation of motion (\ref{eq:smeqmot}) to derive the equation describing the speed jump Eq.~(2) and for  computing the homogenized fracture energy in FIG.~4, where we assumed  Eq.~(\ref{eq:compliance}) with $\lambda_m=0$.

The equation of motion was derived for an elastically homogeneous medium.  However, it faithfully capture the experimentally observed velocity jumps without fitting parameters (see FIG.~2c\&d). Since $\Gstat$ and $g(v)$ are integral properties, they do not change in a discontinuous way between two nearby points. Therefore the effects of elastic heterogeneity at the interface between two materials are minor: $\Gstat$ is approximately constant and cancels out in Eq.~(2) \& (3).
Furthermore, the difference in $g(v)\approx (1-v/\cR)$ for matrix and obstacle material for $v< 0.4 \cR$ is smaller than 8\%, which is within the 10\% variation observed in FIG.~2c\&d. Hence, we use homogenized elasticity when evaluating $g(v)$.

When using the equation of motion to predict the homogenized properties,  we perform an integration over the whole period over which material properties vary (see Eq.~\ref{eq:hom}).  
Thus, homogenized elastic properties can be assumed, as validated by the good agreement between theory and experiments in FIG.~4b.



\section{Trapping and untrapping dynamics}
  
At a material discontinuity, continuity of $\Gstat(\cracklength)$ is ensured, which relates the velocity right before the discontinuity $v^-$ with the velocity immediately after $v^+$ through
\begin{equation}
  \label{eq:sm_equilibrium-at-jump}
 \Gstat= \frac{\fracenergy^-(v^-)}{g(v^-)} =  \frac{\fracenergy^+(v^+)}{g(v^+)}.
\end{equation}
Note that equation~(\ref{eq:sm_equilibrium-at-jump}) does not depend on the direction of propagation but only on material properties. Because a sub-Rayleigh crack has no inertia the trapping and untrapping dynamics is symmetric $\frac{\fracenergy^O(v^O)}{g(v^O)} =  \frac{\fracenergy^M(v^M)}{g(v^M)}$.

\begin{figure}
\includegraphics[width=3.375in]{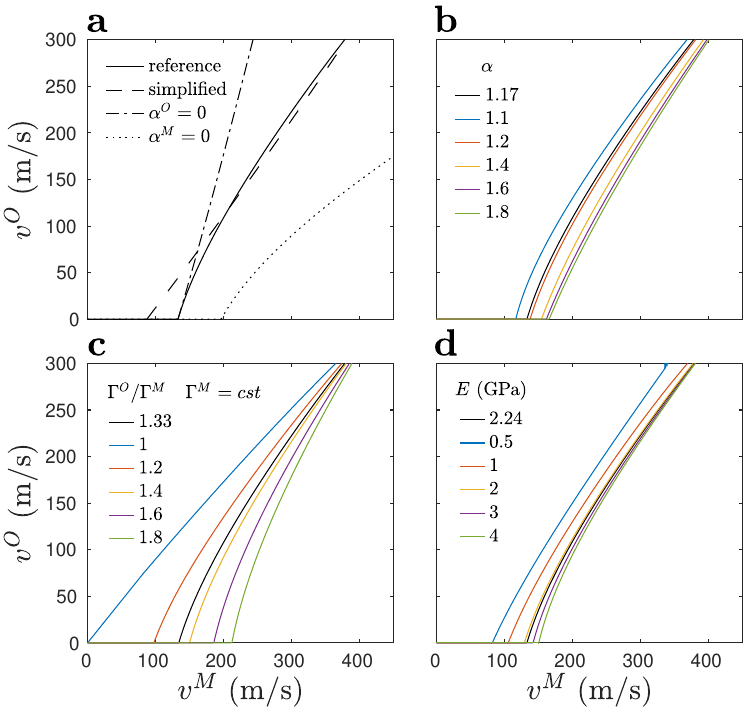}
\caption{\label{fig:smspeed_jump} Trapping and untrapping dynamics assumptions and parametric study. The solid black line is identical throughout the subplots and represents the solution used for comparison with experiments with parameters based on the material characterization (see FIG.~\ref{fig:smcrc}). (a) Reference model (\ref{eq:sm_equilibrium-at-jump}) vs. simplified  model (\ref{eq:sm_simp_jump}) -- dashed line. 
Dash dotted line assumes rate independent obstacle material ($\alpha^O=0$). Dotted line assumes rate independent matrix material ($\alpha^M =0$).
(b, c, d) Parametric study for a wide range of parameter space well beyond the uncertainties of the measurements.}
\end{figure}

A simplified model for the speed jumps at the matrix/obstacle interface is derived based on  the linear version of rate-dependent fracture energy (\ref{eq:simpfrac}).
The velocity jump can be found by solving $\Gstat=\fracenergy^M(v^M)/g(v^M)=\fracenergy^O(v^O)/g(v^O)$ and assuming an average $\gamma=0.2$.
\begin{equation}
  \label{eq:sm_simp_jump}
  v^M-v^O \approx \Delta \fracenergy_0 \frac{1-v^M/\cR}{\gamma +\fracenergy_0^M/\cR}
\end{equation}
The difference between the simplified model  (\ref{eq:sm_simp_jump}) and the reference model (\ref{eq:sm_equilibrium-at-jump}) are shown in FIG.~\ref{fig:smspeed_jump}a. Additionally, we consider the effect of a rate-independent obstacle ($\alpha^O=0$), which causes a decrease in jump amplitude (dash dotted line in FIG.~\ref{fig:smspeed_jump}a). Conversely, a rate-independent matrix  ($\alpha^M=0$) causes a larger velocity jump (dotted line in FIG.~\ref{fig:smspeed_jump}a).

A parametric study is performed on the fracture law parameter $\alpha$ and fracture contrast $\fracenergy^O /\fracenergy^M$ and Young's modulus.
It is observed that $\alpha$ is not very sensitive in the range of uncertainty ($\alpha = 1.17 \pm 10\%$) as shown in FIG.~\ref{fig:smspeed_jump}.b.
Hence, the velocity dependence of the fracture energy is non negligible.
FIG.~\ref{fig:smspeed_jump}.c shows the effect of fracture energy contrast that shifts the graph horizontally as the contrast increases. 
Increasing the Young’s modulus causes a decrease in $v^O$ as depicted in FIG.~\ref{fig:smspeed_jump}.d. The theoretical model in not very sensitive to $\bar E$ in its range of uncertainty ($\bar E = 2.25 \pm 0.2$ GPa). $\bar E$ is the homogenized Young's modulus, which is used to compute $\cR$. $\bar E$ for such a striped composite (with the stripes oriented in the tensile direction) equal to the Reuss bound, which is the harmonic mean of the two moduli: $\bar E = (\frac{1}{2E^M}+ \frac{1}{2E^O})^{-1}$.  
From this parametric study it is concluded that fracture contrast is the most important parameter.

\section{Homogenization of fracture energy}

The equation of motion of a crack (\ref{eq:smeqmot}) is solved numerically assuming $\Gstat(l)=\Gstat_0 e^{-l/\lsys}$ to find $v(l)$ and the results are integrated to compute the homogenized fracture energy  
\begin{equation}\label{eq:hom}
  \homfracenergy =\frac{1}{ \Delta l}\int_{l_i}^{l_i+\Delta l} \fracenergy\left(v(\tilde l)\right) \mathrm{d}\tilde l,
\end{equation}
where $l_i$ is the location of the start of a matrix phase and $\Delta l=w^O+w^M$.
The mean velocity is  $\bar v=\Delta l/\int_{l_i}^{l_i+\Delta l} v^{-1} \mathrm{d}l$.

For vanishing obstacle size with respect to system size $\Delta l/\lsys\rightarrow 0$ we recover a steady-state solution, where $\Gstat$ is constant over a crack advance $\Delta l$. In this case the homogenized fracture energy simply becomes
\begin{equation}
  \homfracenergy =\beta \fracenergy^O(v^O)+ (1-\beta) \fracenergy^M(v^M)
\end{equation}
and the average speed
\begin{equation}
\bar v = (\beta/v^O+(1-\beta)/v^M)^{-1}
\end{equation}
where $w^O=\beta \Delta l$ and $w^M=(1-\beta)\Delta l$.

For further validation, we provide macroscopic fracture energy measurements, based on the applied displacement and the force measured by the load-cell. Essentially, we compute the work of the applied force $F(\delta_m)$ and divide it by the fracture area $A$:
\begin{equation}
\Gamma_{macro}= \frac{1}{A}\int_0^\infty F(\delta_m)\dd \delta_m
\end{equation}

\begin{figure}[h]
    \centering
    \includegraphics[width=3.375in]{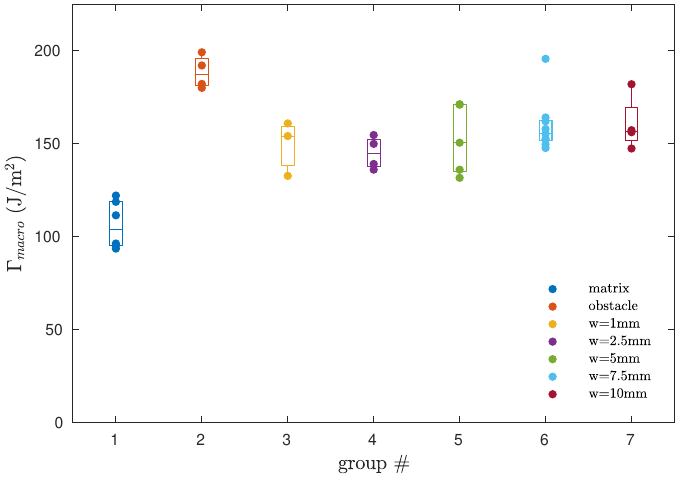}
    \caption{Macroscopic measurement of fracture energy.  The average macroscopic fracture energy of matrix is 105J/m$^2$, of obstacle is 190J/m$^2$, and of the heterogeneous samples is  155J/m$^2$. }
    \label{fig:macro_frac_energy}
\end{figure}

These measurements show that the macroscopic fracture energy of the heterogeneous samples corresponds to the average of the obstacle and matrix fracture energy and are shown in Fig.~\ref{fig:macro_frac_energy}.

\section{Digital Image Correlation analysis}

Digital Image Correlation consists in providing an estimate measurement of the displacement field $\bm{u}$ at a given time $t$ based on correlating the deformed image $f(\bm{x},t)$ with its reference undeformed image $f(\bm{x},t=0)$, where $\bm{x}$ is the spatial coordinate.

The basic principle is the conservation of gray level
\begin{equation}
  f(\bm{x},t=0) = f(\bm{x}+\bm{u},t),
\end{equation}
from which a minimization problem can be set up:
\begin{equation}
  \label{eq:dic}
\bm{u}= \mathrm{argmin} \int_{A} R(\bm{u})^2 \mathrm{d} \bm{x},
\end{equation}
where $R(\bm{u})=|{f(\bm{x},t=0)-f(\bm{x}+\bm{u},t)}|$ is the residual and $A$ the region of interest.
In order to find the displacement field $\bm{u}$ and inverse problem needs to be solved.

There are multiple approaches for solving the inverse problem, the simplest of which being subdividing the domain in sub-domains $A_i$ and finding $\bm u_i$ by cross correlation. In the current study different approaches have been used: a global approach and an integrated approach.

\emph{Global Approach:}
A global digital image correlation \cite{roux_stress_2006} involves resolving $\bm u$ by minimizing $R(\bm{u})$ over the whole region of interest.
$\bm{u}$ is approximated by
\begin{equation}
\bm{u}(\bm{x})=\sum \bm{u}_i \phi_i(\bm{x})  
\end{equation}
where $\phi_i(\bm{x})$ are basis functions and $\bm{u}_i$ the displacement of each degree of freedom.
In order to ensure algorithmic stability, a regularization term which penalizes high gradients in $\bm u$ is introduced. 
\begin{equation}
\bm{u} = \mathrm{argmin} \int_{A} \left( R(\bm{u})^2 \mathrm{d} + \alpha ||\nabla \bm{u} ||^2\right) \mathrm{d} \bm{x}
\end{equation}

\emph{Integrated Approach:} 
The Integrated Digital Image Correlation method  \cite{roux_stress_2006,grabois_validation_2018} uses the analytical solution for a straight crack in an infinite linear elastic medium, known as the  Williams eigenfunction expansion $r^{n/2}\psi_n(\theta)$, as basis for the trial displacement field
$u=\sum_n a^I_n r^{n/2}\psi^I_n(\theta) + \sum_n a^{II}_n r^{n/2} \psi^{II}_n(\theta)$
\cite{williams_stress_1956}, where $r$ is the distance from the crack tip and $\theta$ the angle with respect to the propagation direction. The superscripts I and II stand for the mode of fracture, I being symmetric with respect to the crack plane and II anti-symmetric. Note that the terms for $n=0$ are rigid body motions, $n=1$ represents the singular stress fields and discontinuous displacements across the crack and are related to the stress intensity factor 
\begin{equation}
a^I_1 =\frac{\K^I}{ \mu \sqrt{2\pi}} , \quad
a^{II}_1 = \frac{\K^{II}}{ \mu \sqrt{2\pi}}
\end{equation}
which in turns is related to the fracture energy by Eq.~(\ref{eq:dynamicG}).
Higher order terms are governed by boundary conditions.
Crack velocity is slow enough $v<0.4\cR$ that the velocity dependence of the angular functions $\psi_n(\theta,v)$ (see Eq.~\ref{eq:disp_dyn}) can be neglected \cite{svetlizky_classical_2014}.

The near-tip displacement field can be represented by the first 10 terms of the expansion, which greatly reduces the number of unknowns to be solved for.
Thus, there is no need for penalizing high gradients. Also, it automatically accounts for the displacement discontinuity caused by the presence of the crack.

The crack tip position is determined by considering the location along the crack path where the first supersingular term of the Williams expansion $n=-1$ vanishes. This allows for precise measurement of the crack tip position.
The amplitudes $a^I_n$ and $a^{II}_n$ are obtained by solving  (\ref{eq:dic}).

\emph{Effect of elastic heterogeneity on IDIC}
The William's expansion assumes that the material elastic properties are homogeneous. The matrix and obstacle material used in the heterogeneous samples differ in Young's modulus by 30\%. The IDIC analysis is used to determine the crack tip position and fracture energy.
The crack tip position is directly determined by the $u\sim r^{1/2}$ term in the Williams' expansion, which is the dominant term in the displacement field at the tip. Since, the position of the tip does not depend on the amplitude of this term, the elastic heterogeneity does not affect the precision of determining the crack-tip position.

Regarding the fracture energy, we validated the viability of using IDIC on the heterogeneous samples by measurements on homogeneous samples. Our data suggest that effects of elastic heterogeneity of 30\% are within the range of variation observed in the homogeneous samples (see FIG.~\ref{fig:smcrc}). This can be explained by a homogenization argument, where elasticity near the crack-tip are effectively homogenized within the near tip region where the displacement field $u\sim r^{1/2}$. The size of this region is $\sim1$cm which is larger than the size of the heterogeneity.
The interplay between the size of the K-dominant region and the size of the heterogeneity is important. The strain at short distance could follow a local K-field, while at large distance it would behave as a homogenized K-field. However, if the size of the K-dominant region was small, a homogenized K-field could not be established. We will investigate this cross-over in future work, on samples with larger elastic contrast.

\emph{Speckle Pattern:}
For correlating reference and deformed images over the region of interest surrounding the crack tip \cite{grabois_validation_2018}, a speckle pattern was applied on the surface of the specimen using aerosol paint. First, a homogeneous white coating is applied, followed by a black speckle sprayed from a larger distance. Spline interpolation of the gray-levels across pixels allowed for sub-pixel displacement measurements. The auto-correlation length of the speckle pattern corresponds to 4-6 pixels. The pixel size is $45~\mu\text{m}$.

\begin{figure*}
    \centering
    \includegraphics[width=5in]{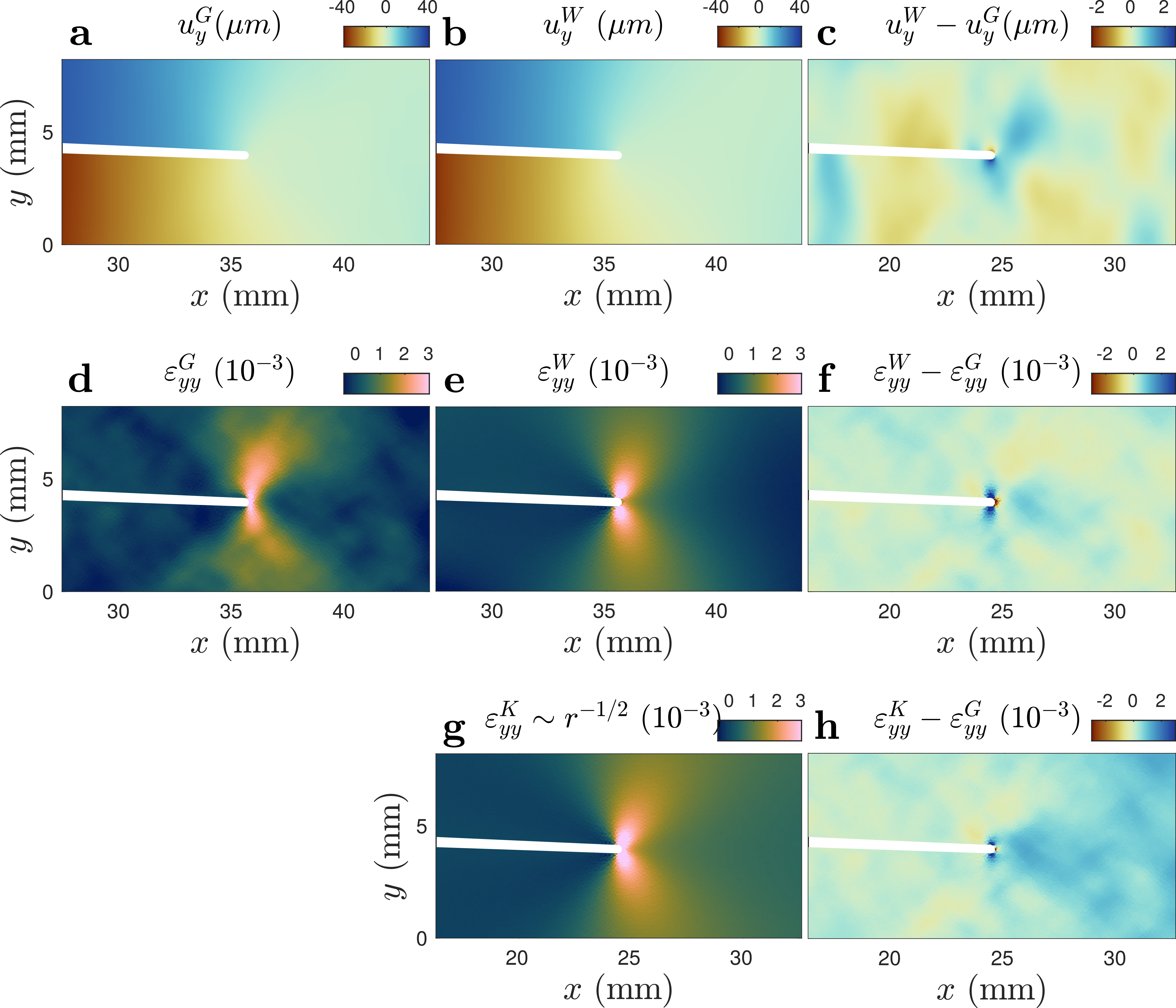}
    \caption{Digital Image Correlation results. (a)~Displacement field $u_y$ measured using the global approach. (b)~$u_y$ using integrated approach. (c)~Difference in measured displacement between the two methods.
    (d)~Strain field $\varepsilon_{yy}=\partial_y u_y$ computed from the results of the global approach. 
    (e)~$\varepsilon_{yy}$ from the integrated approach.
    (f)~Difference in strain between the two methods.
     (g)~Singular component of $\varepsilon_{yy}$ from the integrated approach.
    (h)~Difference in strain between the global method and the the singular component from the integrated method.}
    \label{fig:dic_fields}
\end{figure*}


\onecolumngrid

\section{Williams eigenfunctions expansion}

Consider a polar coordinate system with origin at the crack tip.
\begin{equation}
 x - x_\text{tip} + i y= r e^{i\theta}
\end{equation}
The displacement field $u = u_x + i u_y$ is a linear combination of the eigenfunctions
\begin{equation}
  u = \sum_n a^I_n  r^{n/2} \psi_n^I(\theta) + \sum_n a^{II}_n  r^{n/2} \psi^{II}_n(\theta)
\end{equation}
where
  \begin{alignat}{3}
  &\psi^{I}_n(\theta) &&=    \frac{1}{2}\Big(    &&\kappa e^{i\theta \frac{n}{2}} - \frac{n}{2} e^{i\theta(2-\frac{n}{2})} + (-1)^n e^{-i\theta \frac{n}{2}} +\frac{n}{2}  e^{-i \theta \frac{n}{2}}\Big)\\
  &\psi^{II}_n(\theta) &&= \frac{i}{2}\Big(&&\kappa e^{i\theta \frac{n}{2}} + \frac{n}{2} e^{i\theta(2-\frac{n}{2})} + (-1)^n e^{-i\theta \frac{n}{2}} -\frac{n}{2}  e^{-i \theta \frac{n}{2}} \Big)
\end{alignat}
where, for plane-stress conditions,
\begin{equation}
  \kappa = \frac{3-\nu}{1+\nu},
\end{equation}
with $\nu$ the Poisson's ratio.

\subsection{Near tip fields for in-plane dynamic cracks}

In the following section we report the solution of  the asymptotic near tip fields of a dynamic mode I crack as derived following the procedure in  \citep{freund_dynamic_1990}.

For a dynamic crack the polar coordinates $r,\theta$ are related to the velocity dependent coordinates $r_{d,s},\theta_{d,s}$ according to

\begin{alignat}{2}
    &\gamma_d =r_d/r = \sqrt{1-(v\sin\theta/c_d)^2} \quad &&\gamma_s =r_s/r=\sqrt{1-(v\sin\theta/c_s)^2}\\
    &\tan\theta_d=\alpha_d\tan\theta,\quad &&\tan\theta_s=\alpha_s\tan\theta,
\end{alignat}
where $\alpha_d^2\equiv 1-v^2/c_d^2$ and $\alpha_s^2\equiv 1-v^2/c_s^2$.

The mode I near tip stress field is
\begin{equation}
    \sigma_{ij}(r,\theta) = \sum_{n} \mu b_{n} r^{n/2}\Sigma^I_{ij,n}(\theta,v)
\end{equation}
where the asymptotic angular functions $\Sigma^I_{ij,-1}$ are
\begin{alignat}{2}
    &\Sigma^I_{yy}(\theta,v) &&= - \frac{1}{D(v)}\left[(1+\alpha_s^2)^2 \gamma_d^{-1/2} \cos(\theta_d/2)-4\alpha_s \alpha_d \gamma_s^{-1/2}\cos(\theta_s/2)\right]\\
    &\Sigma^I_{xy}(\theta,v) &&=  \frac{2\alpha_d(1+\alpha_s)^2}{D(v)}\left[ \gamma_d^{-1/2} \sin(\theta_d/2)- \gamma_s^{-1/2}\sin(\theta_s/2)\right]
\end{alignat}
where $b_{-1}$ is related to the stress intensity factor by
\begin{equation}
b_{-1}=\frac{K_I}{\mu \sqrt{2\pi}}
\end{equation}
where
\begin{equation}
    D(v)=4\alpha_s \alpha_d - (1+\alpha_s^2)^2
\end{equation}
is the Rayleigh function.

Similarly the mode I displacement field is
\begin{equation}\label{eq:disp_dyn}
    u_i(r,\theta) = \sum_n b_n r^{1+n/2}\psi^I_{i,n}(\theta,v)
\end{equation}
where the asymptotic angular functions $\psi_{i,-1}(\theta,v)$
\begin{alignat}{2}
&\psi^I_{x,-1}(\theta,v) &&= 2\frac{1}{D(v)}\left[(1+\alpha_s^2)\gamma_d^{1/2}\cos(\theta_d/2)-2\alpha_s\alpha_d \gamma_s^{1/2}\cos(\theta_s/2)\right]\\
&\psi^I_{y,-1}(\theta,v) &&= - 2 \frac{\alpha_d}{D(v)}\left[(1+\alpha_s^2)\gamma_d^{1/2}\sin(\theta_d/2)-2 \gamma_s^{1/2}\sin(\theta_s/2)\right]
\end{alignat}

The mode I stress angular functions (see \cite{svetlizky_classical_2014} for mode II) consider odd terms $n=-1,1,3,5,...$
\begin{alignat}{2}
    &\Sigma^I_{yy,n}(\theta,v) &&= -\frac{1}{D(v)}\left[(1+\alpha_s^2)^2 \gamma_d^{n/2} \cos(n\theta_d/2)-4\alpha_s \alpha_d \gamma_s^{n/2}\cos(n\theta_s/2)\right]\\
    &\Sigma^I_{xy,n}(\theta,v) &&= -\frac{2\alpha_d(1+\alpha_s)^2}{D(v)} \left[\gamma_d^{n/2} \sin(n\theta_d/2)- \gamma_s^{n/2}\sin(n\theta_s/2)\right]
\end{alignat}

\iffalse
\bibliography{MyLibrary.bib}
\else

\fi